\documentclass[conference,compsoc]{IEEEtran}

\usepackage{dingbat}
\usepackage{listings}
\usepackage{color}
\usepackage{makecell}
\usepackage{array}
\usepackage{xcolor}
\usepackage{multirow}
\usepackage{colortbl}
\usepackage{enumitem}
\usepackage{adjustbox}
\usepackage{amsfonts}
\usepackage{subcaption}
\usepackage[linesnumbered,ruled,vlined]{algorithm2e}
\SetAlgorithmName{Algorithm}{Algorithm}{Algorithm}

\SetAlgoVlined
\SetAlgoLined
\SetKwInput{KwInput}{Input}%
\SetKwInput{KwOutput}{Output}%
\SetKwProg{Proc}{Procedure}{}{}

\DontPrintSemicolon
\usepackage{graphicx}
\usepackage[most]{tcolorbox}
\usepackage{tabularx}

\usepackage{algpseudocode}
\usepackage{url}
\usepackage{float}
\usepackage[colorlinks=true, linkcolor=blue, citecolor=violet, urlcolor=blue]{hyperref}
\usepackage{caption}
\usepackage{adjustbox}

\algtext*{EndProcedure}%

\definecolor{mypurple}{HTML}{67379A}
\definecolor{myblue}{HTML}{93AAD8}
\definecolor{myred}{HTML}{EB3323}
\definecolor{myyellow}{RGB}{204,153,0}
\definecolor{mygreen}{RGB}{0,102,0}
\definecolor{greenbg}{rgb}{0.9, 1, 0.9}
\definecolor{scfiNavy}{HTML}{1F4E79}
\definecolor{scfiBrick}{HTML}{A8432C}
\definecolor{scfiGreen}{HTML}{4F7F5A}

\usepackage{acro}
\DeclareAcronym{llm}{
  short = LLM,
  long  = large language model
}
\DeclareAcronym{skillcfg}{
  short = SkillCFG,
  long  = Skill Control-Flow Graph
}
\DeclareAcronym{cfi}{
  short = CFI,
  long  = control-flow integrity
}
\DeclareAcronym{cfg}{
  short = CFG,
  long  = control-flow graph
}

\lstdefinestyle{mystyle}{
    backgroundcolor=\color{white},
    basicstyle=\ttfamily\footnotesize,
    breakatwhitespace=false,
    breaklines=true,
    captionpos=b,
    keepspaces=true,
    numbers=left,
    numbersep=5pt,
    showspaces=false,
    showstringspaces=false,
    showtabs=false,
    tabsize=2,
    numberstyle=\tiny\color{black},
    framexleftmargin=5mm, frame=single, framerule=0pt,
    xleftmargin=5mm,
    xrightmargin=3mm,
    framesep=5mm,
    rulecolor=\color{black},
    postbreak=\mbox{\textcolor{red}{$\hookrightarrow$}\space},
}
\usepackage{amsthm}
\theoremstyle{definition}
\newtheorem{defn}{Definition} 
\theoremstyle{plain}

\newtheorem{prop}{Proposition}

\makeatletter
\@ifundefined{prop}{\newtheorem{prop}{Proposition}}{}
\makeatother

\newcommand{\tool}{SkillCFI\xspace}

\usepackage{pifont}

\usepackage{tikz}
\usepackage{amsmath}

\definecolor{circledblue}{RGB}{46,95,127}

\definecolor{mappingblue}{RGB}{46,116,181}
\definecolor{mappinggrey}{RGB}{230,230,230}



\usepackage{booktabs}

\usepackage{fvextra}

\DefineVerbatimEnvironment{myverbatim}{Verbatim}{
  breaklines=true,
  breakanywhere=true
}

\makeatletter
\newcommand{\removelatexerror}{\let\@latex@error\@gobble}
\makeatother

\makeatletter
\def\tcb@cnt@datalistautorefname{Listing}
\makeatother

\usepackage{color}
\usepackage{soul}
\usepackage{pifont}
\usepackage{lipsum}%
\usepackage{wasysym}
\SetSymbolFont{wasy}{bold}{U}{wasy}{m}{n}
\newtcolorbox[auto counter]{datalist}[2][]{%
  enhanced,
  breakable,%
  float,
  floatplacement=t,
  fonttitle=\bfseries\small,
  fontupper=\small,
  colback=gray!10,%
  colframe=black,%
  fonttitle=\bfseries,%
  title={Listing \thetcbcounter. #2},%
  title after break={Listing \thetcbcounter. #2 (continued)},%
  rounded corners,%
  boxrule=0.5pt,%
  top=2pt, bottom=2pt, left=2pt, right=2pt,%
before=\par\vspace{5pt}\noindent,%
  after=\par,%
  before upper={\setlength{\parindent}{0pt}},%
  width=\linewidth,%
  #1%
}

\newcommand{\cnone}{\mbox{\textsc{No-Gate}}\xspace}
\newcommand{\callow}{\mbox{\textsc{Allowlist}}\xspace}
\newcommand{\cnorel}{\mbox{\textsc{No-Rel}}\xspace}
\newcommand{\cfull}{\mbox{\textsc{SkillCFI}}\xspace}
\newcommand{\fbare}{\mbox{\textsc{Bare}}\xspace}
\newcommand{\fbasic}{\mbox{\textsc{Basic}}\xspace}
\newcommand{\fguid}{\mbox{\textsc{Next-Step}}\xspace}

\newtcolorbox{rqanswerbox}{%
  enhanced, colback=scfiNavy!4, colframe=scfiNavy!45, boxrule=0.4pt, arc=1pt,
  left=4pt, right=4pt, top=3pt, bottom=3pt, fontupper=\small,
  before skip=6pt, after skip=6pt, width=\linewidth}

\usepackage{nameref}%

\SetKwIF{If}{ElseIf}{Else}{if}{then}{else if}{else}{endif}
\SetKwFor{For}{for}{do}{endfor}
\SetKwFor{While}{while}{do}{endw}
\SetKw{Return}{return}
\SetKw{Require}{require}
\SetKwInput{KwState}{Monitor State}

\makeatletter
\@ifundefined{thm}{%
  \theoremstyle{plain}%
  \newtheorem{thm}{Theorem}%
}{}
\makeatother

\ifCLASSOPTIONcompsoc
  \usepackage[nocompress]{cite}
\else
  \usepackage{cite}
\fi

\ifCLASSINFOpdf
\else
\fi

\begin{document}
\title{Guarding the Kingdom of Skill-Based Agents in the AI Dynasty of Computation: Control-Flow Integrity for Agent Skills}

\author{\IEEEauthorblockN{Bonan Ruan\textsuperscript{1}\quad
Yuhao Sun\textsuperscript{2}\quad
Jiahao Liu\textsuperscript{1}\quad
Qixiao Lin\textsuperscript{2}\quad
Jian Mao\textsuperscript{2}\quad
Zhenkai Liang\textsuperscript{1}}
\IEEEauthorblockA{\textsuperscript{1}National University of Singapore\qquad
\textsuperscript{2}Beihang University}}

\maketitle

\begin{abstract}
Control-flow integrity (CFI) confines a program to a control-flow graph fixed ahead of time.
Procedural agent skills move this control flow into prompt text, and a large language model (LLM) agent decides at runtime whether each step runs in its reasoning, a tool call, or a script.
The procedure's \emph{control skeleton} survives, but its enforceability depends on where each step runs.
Many violations, such as overwriting a file before validation, use only operations the skill requires, so permission allowlists admit them.
A deviating or prompt-injected model can also evade a monitor by \emph{regime migration}, for example by claiming that a check passed.

We present SkillCFI, which carries CFI over to skill-based agents.
Its compiler confines LLM interpretation to compile time and derives the agent's instructions and an executable contract from one skill.
By a formal enforceability criterion, it enforces obligations whose evidence crosses the tool boundary and rejects a skill whose mandatory obligation leaves nothing to check.
A deterministic runtime enforcer admits a tool request only on trusted evidence and refuses unmediated channels such as inline scripts, so admitted traces follow the contract whatever the model proposes.

On eight self-written skills, no compiled contract permits more than its source under 24 source mutations.
In 770 agent sessions, SkillCFI blocks before dispatch all 240 such violations, which an allowlist admits, and normal tasks complete nearly as often as without it.
Listing the next legal steps in a refusal raises recovery from 21\% to 65\%.
Checking takes about 2\% of task time, and a compact command form cuts session slowdown from $1.32\times$ to $1.08\times$.
None of 53 public skills is accepted as written.
The compiler's diagnostics point to interfaces rather than workflows and guided seven into enforceable contracts.
\end{abstract}

\IEEEpeerreviewmaketitle

\acresetall
\section{Introduction}
\label{sec:introduction}

Software is the embodiment of a computation task, traditionally represented by algorithms and implemented by programs.
The evolution of software systems is the continual redistribution of computation across the changing components of the computing platform.
Each technological era sends forth its own waves of new architectures, languages, and computing paradigms.
They rise, compete, and eventually recede.
Some are swept away by the next tide, while others survive by being absorbed into newer forms.
Beneath these waves, the underlying computation endures as an abstract algorithm.
Periods of technical stability tend to create clear partitions, abstractions, and principled boundaries.
As systems mature, accumulated demands for new features eventually erode these boundaries, so that concepts become entangled and system principles are neglected.
Through this evolution, while the overall computation stays unchanged, it is redistributed across emerging components and partitions.

\Ac{llm}-based agents exemplify this redistribution of computation and the need to re-establish protection and boundaries.
They execute multi-step workflows on behalf of users~\cite{yao2022synergizing,DBLP:conf/icml/WangCY0L0J24,tang2026devops} by calling external tools, such as file operations and shell commands.
Increasingly they follow \emph{Agent Skills}, reusable procedures written in natural-language \texttt{SKILL.md} documents~\cite{agentskillsOverviewAgent,jiang_sok_2026}.
A skill tells the agent how to carry out one kind of task.
It names the steps, their order, the checks that must pass, and the points where the user decides.
A document-editing skill, for example, may require the agent to validate a revised draft before delivering it and to ask the user before overwriting the source file.
These are what a program encodes as control flow.
Skills move this logic from code into prose and hand its execution to a model.

Nothing holds the agent to this procedure.
The model reads the skill as part of its prompt, next to the task and the content it processes, and decides by itself what to do next.
It may skip a check, take the wrong branch, or report a step as done that never ran.
Such deviations happen even without an attacker, because models follow long procedures unreliably~\cite{yao2025tau}.
On compliance-critical skills, agents that execute \texttt{SKILL.md} directly satisfy only 26.2--54.0\% of the prescribed requirements on average~\cite{dantanarayana2026sigil}.
Agents also claim completion 28.7--37.9 percentage points more often than official evaluators confirm it~\cite{li2026holds}.
An attacker can also trigger such deviations deliberately through indirect prompt injection, for example with text in a task file that claims a check has passed~\cite{greshake2023not,debenedetti2024agentdojo}.
Crucially, many of these violations use only operations that the skill itself requires.
Overwriting the source is a lawful step once the draft is validated and the user agrees, so restricting which tools the agent may call~\cite{shi_progent_2025,wu2026skillscope,pan2026skillguard} does not prevent them.

\begin{figure}[t]
  \centering
  \includegraphics[width=\linewidth]{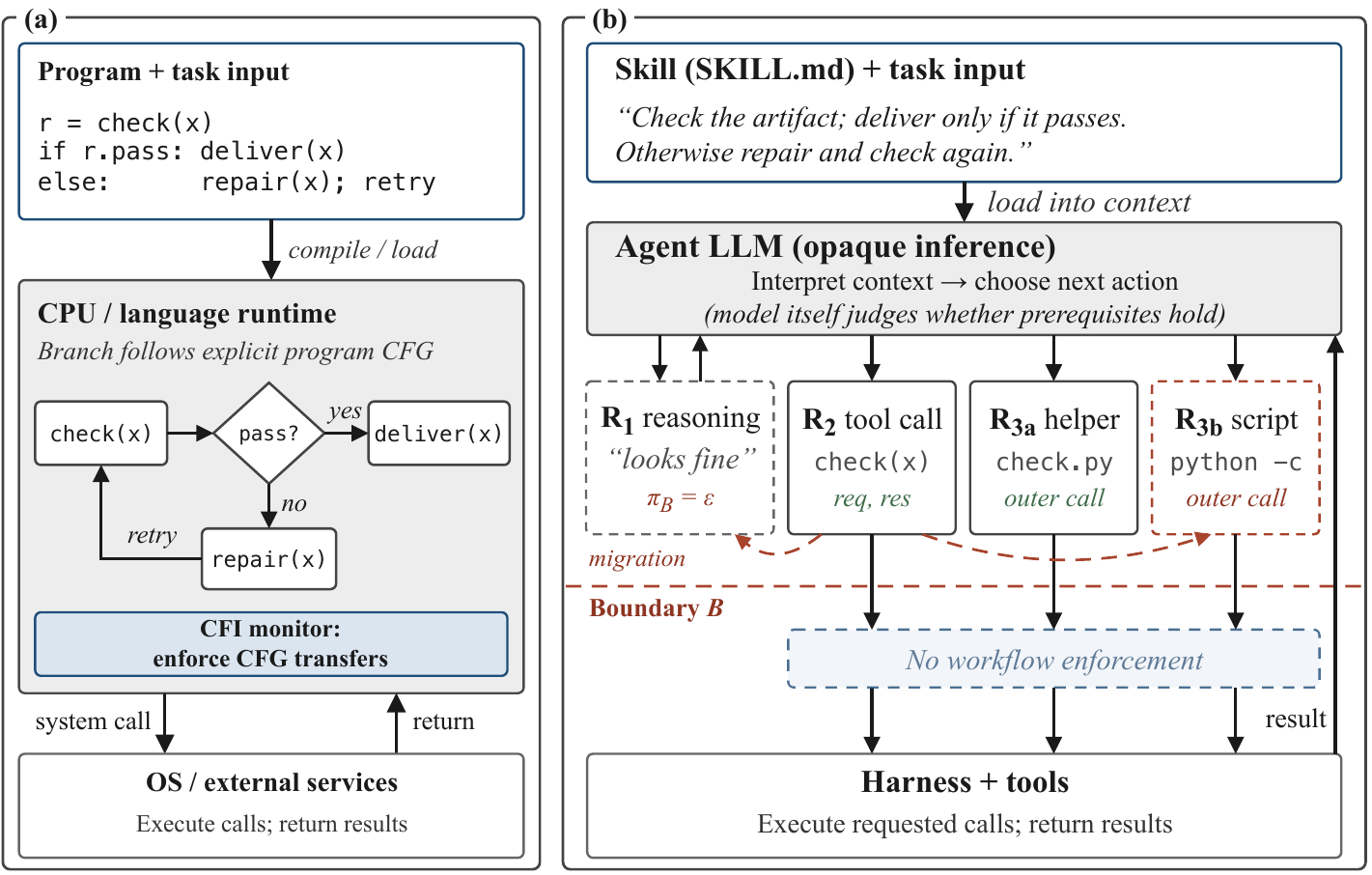}
  \caption{One check-and-repair procedure as a program and as a skill. (a)~An independent monitor checks program execution against an explicit control-flow graph. (b)~The model decides where each step runs. Relative to the tool-call boundary $B$, a step in reasoning ($R_1$) leaves no event, a tool call ($R_2$) exposes its request and result, and a helper ($R_{3\mathrm{a}}$) or generated script ($R_{3\mathrm{b}}$) exposes only its outer call.}
  \label{fig:execution-paradigms}
\end{figure}

Each of these deviations runs a step where the procedure forbids it, so it violates the skill's control flow.
In programs, such violations have a classical defense.
As \autoref{fig:execution-paradigms}(a) shows, a program encodes the control structure of its task as an explicit \ac{cfg}, and \ac{cfi}~\cite{abadi_control-flow_2009} uses an independent monitor to confine execution to that graph.
A skill-based agent has no explicit graph and no such monitor.
Its procedure exists only as prompt text, and the agent decides at runtime whether to compute each step in its own reasoning, through a tool call, or inside a script it invokes or generates~\cite{DBLP:conf/icml/WangCY0L0J24}.
The part a program would encode in its \ac{cfg} is the skill's \emph{control skeleton}.
It consists of \emph{procedural obligations}, such as mandatory steps, prerequisites, outcome-dependent branches, and repetition bounds, while \emph{semantic work}, such as revising a document, remains with the model.
No component outside the model checks a request against the skeleton before it executes (\emph{No workflow enforcement} in \autoref{fig:execution-paradigms}(b)).
The model that proposes an action also decides whether its prerequisites hold, so the boundary between proposing and authorizing has eroded.

Recent skill defenses aim at reliability rather than at a deviating or manipulated agent.
SIGIL takes execution away from the agent~\cite{dantanarayana2026sigil}, and SkillSentry checks the agent through hooks but credits progress from its own proposals~\cite{lu2026skillsentry}.
History-aware monitors~\cite{wang2026agentspec,palumbo_policy_2026,li2026vigil,li2026holds} can express the required orderings.
Their policies, however, come from hand-written rules or \ac{llm} translations that are separate from the agent's instructions, so the calls the agent is taught may not be the calls the monitor accepts.
It remains open which skill obligations a monitor can enforce when the agent chooses where each step runs, and how to enforce them while the agent keeps control.

Our key observation is that a skill's control skeleton does not depend on where its steps execute, but the enforceability of each obligation does.
The skeleton is thus a fixed graph like a program's \ac{cfg}, and \ac{cfi} can enforce it to the extent that a monitor can observe its steps.
In current agent hosts, a step leaves the model at a tool call.
A host monitor sees only this mediation boundary $B$, which carries tool requests before dispatch, their results, and the user's answers to questions asked through the host.
Each step runs in one of three execution regimes relative to $B$ (\autoref{fig:execution-paradigms}(b)), namely the model's reasoning ($R_1$), a mediated tool call ($R_2$), or a helper or generated program ($R_3$).
A monitor can enforce an obligation only under two conditions.
Its protected effect must cross $B$ before it executes, and its prerequisites must be established by trusted results at $B$.
Because the model chooses the regime at runtime, it can evade the monitor by moving a protected step out of $R_2$, for example into an inline script.
To prevent such \emph{regime migration}, every channel that can produce a protected effect must be mediated.

Carrying \ac{cfi} over to skills thus needs an explicit graph and a protected monitor, which raises three challenges.
\textbf{C1: Making the skeleton explicit without changing it.}
Extracting the skeleton from prose requires an \ac{llm}, which may omit a branch, invent a bound, or produce an ambiguous result.
The extracted contract must neither weaken a mandatory requirement nor remove a permitted alternative.
\textbf{C2: Sharing one interface between instructions and monitor.}
The agent must be taught exactly the call forms that the monitor recognizes.
When one operation recurs, the monitor must also resolve which step a request belongs to.
\textbf{C3: Enforcing without taking control.}
The monitor must stay sound whatever the model proposes, refuse a request without changing the workflow state, and leave the choice among legal paths to the model.

We present \tool, which carries \ac{cfi} over to skill-based agents.
Its offline compiler confines \ac{llm} interpretation to compile time, where an \ac{llm} proposes the interfaces and a workflow specification.
It marks each source obligation as \texttt{enforce}, \texttt{guidance}, or \texttt{unsupported}, and stops with a diagnostic when a mandatory one is unsupported.
A deterministic verifier rejects specifications with overlapping interfaces, unhandled outcomes, unavailable evidence, or ambiguous step matches (C1).
From the accepted specification, the compiler generates the agent's instructions and an executable contract, so the agent and the monitor share the same call forms (C2).
At runtime, the enforcer consults no model and keeps workflow state outside the model's context.
It checks each request before dispatch, advances only on the results of calls it admitted, and refuses unmediated channels such as inline scripts.
A refusal leaves this state unchanged and returns the next legal steps (C3).

This paper makes the following contributions.

\begin{itemize}[leftmargin=*]
    \item \textbf{Skills as redistributed control flow.}
    We analyze skill security in the context of software evolution, in which each new platform redistributes the same computation and a skill moves the judgments and transfers of its control skeleton into the model (\autoref{sec:background}).
    Procedure violations are thus control-flow violations, which \ac{cfi} addresses (\autoref{sec:threat-security}).

    \item \textbf{Enforceability, compilation, and enforcement.}
    We model where an agent runs each skill step relative to the tool-call boundary and identify regime migration (\autoref{sec:execution-model}).
    Building on classical enforceability results~\cite{schneider2000enforceable,basin2013enforceable,ligatti2010theory}, we derive which obligations a boundary monitor can enforce, which stay as guidance, and which force rejection, and we state two principles that close migration.
    We design a propose-and-check compiler and a runtime enforcer whose admitted traces follow the contract (\autoref{sec:approach}), and implement them for the OpenCode agent~\cite{opencodePlugins}.

    \item \textbf{Evaluation.}
    We evaluate \tool on eight self-written skills with 770 agent sessions and on 53 public skills (\autoref{sec:evaluation}).
    It blocked before dispatch all 240 violations that use only the skill's own operations, while an allowlist let all 80 such violations in its sessions take effect.
    An offline analysis shows that a monitor that admits all legal requests must still admit all 160 violations executed under the baselines if it decides on one request at a time, and 59 of them if it sees requests but not their results.
    Normal tasks complete nearly as often as without enforcement, and checking takes about 2\% of task time.
    No public skill is accepted as written, and the compiler's diagnostics guided seven selected ones into enforceable contracts.
\end{itemize}

\section{Insights and Motivation}
\label{sec:background}

\subsection{Software Architecture Evolution \& Invariants}
\label{sec:redistribution}
\label{sec:llm-agents}

A computation task is an abstract algorithm.
The algorithm fixes the steps to run, the judgments that choose among branches, how often a step may repeat, which object each step acts on, and when the task ends.
We call this structure the task's \emph{control skeleton}.
It is a virtual \ac{cfg} that exists whether or not any artifact writes it down, and every correct execution of the task follows one of its paths.

An implementation distributes the algorithm over the components of its platform.
For example, a compiled program places the whole skeleton into its binary code, where conditional branches realize the judgments and jumps realize the transfers.
Later platforms move parts of the skeleton into other components, such as an interpreter, a remote service, or a configuration.
A redistribution changes where these parts run, but not what the task requires.
The skeleton, the virtual \ac{cfg} of the task, remains the invariant.

A skill-based agent, a recent software architecture, combines an \ac{llm}, a harness, and external tools.
The \emph{harness} is the host software that assembles the model's context, dispatches the model's tool requests, and returns their results as observations.
In the Agent Skills format, a \emph{skill} is a directory with a \texttt{SKILL.md} file, whose metadata says when the skill applies and whose body gives the procedure~\cite{agentskillsOverviewAgent,jiang_sok_2026}.
It may also bundle helper scripts, references, and templates.
The \emph{task input} gives one run's parameters, such as target files or preferences.

This architecture redistributes the computation task.
The task's control skeleton moves from code into the prose of \texttt{SKILL.md}.
At runtime, the model reads this prose and carries out each step in one of three places.
It can perform the step with reasoning, issue it as a tool call that the harness dispatches, or hand it to a program, either a helper bundled with the skill or a script that the model writes.
The judgments and transfers stay in the model.
The model decides whether a check passed, which branch to take, whether another repair is allowed, and when the task ends.

\begin{figure}[t]
  \centering
  \includegraphics[width=0.95\linewidth]{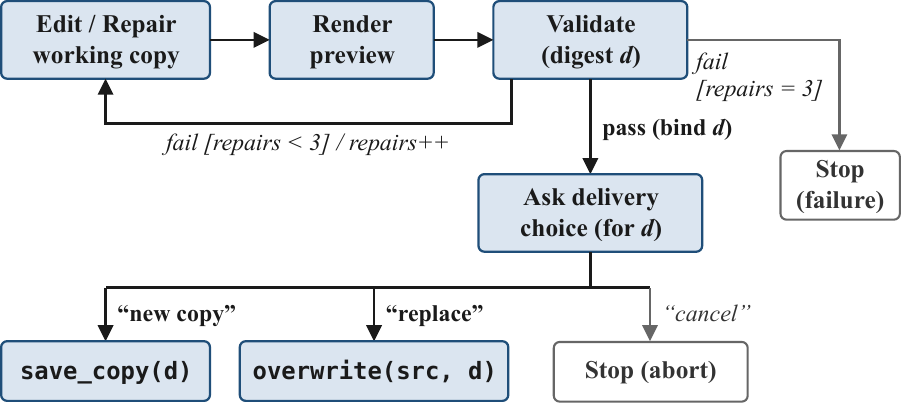}
  \caption{The document-editing skill, which revises the abstract of \texttt{source.md}.}
  \label{fig:document-editing-motivation}
\end{figure}

\noindent\textbf{Running example.}
\autoref{fig:document-editing-motivation} shows the control skeleton of a document-editing skill.
The agent must edit and render a working copy, pass validation with at most three repairs, ask the user how to deliver the validated draft, and deliver exactly that draft, identified by its digest $d$.
The skeleton has a judgment on the validation outcome, a bounded repair loop, a branch on the user's choice, and a binding of the delivered content to $d$.

\subsection{Control-Flow Integrity}
\label{sec:background-cfi}

\Ac{cfi} is the classical way to hold a program to its control skeleton.
It requires that execution follow a \ac{cfg} determined ahead of time~\cite{abadi_control-flow_2009}.
Most transfers of a program are fixed in its code, and the ones at risk are computed transfers, such as indirect calls and returns, whose targets come from data an attacker may corrupt.
\ac{cfi} locates each computed transfer and restricts it to valid targets before control reaches them.

\ac{cfi} is a policy, and it can be enforced in more than one way.
The original mechanism is an inline reference monitor that instruments each computed transfer.
Whatever the mechanism, enforcement needs three things.
(i) \textbf{A graph fixed ahead of time.}
The \ac{cfg} is derived from the program before it runs.
(ii) \textbf{A check at every computed transfer.}
The check runs before the transfer takes effect, and the program cannot skip it.
(iii) \textbf{A protected monitor.}
The program cannot alter the checks or the state they rely on.
\ac{cfi} enforces adherence to the graph, so an attack that stays within permitted control transfers is out of its scope~\cite{carlini2015control}.

This paper carries the \ac{cfi} policy over to agent skills.
A skill has a control skeleton as a program does, but none of the three requirements holds as given.
For (i), its graph exists only as prose.
For (ii), its transfer points are opaque.
A program's computed transfers are instructions that a monitor can instrument.
A model's transfers happen inside inference, and the text in which it explains them is only a claim.
Moreover, every transfer is computed dynamically, since the model derives the next step from a context mixed with untrusted task content.
For (iii), workflow state, such as whether validation passed, lives only in that context.

A check therefore cannot sit at the transfer itself.
Security mechanisms place a reference monitor on a boundary that untrusted computation must cross, and check what crosses it against a model of the intended behavior~\cite{anderson1972computer,saltzer1975protection}.
When part of the skeleton moves into a component that a monitor cannot inspect, the check must move to where that part becomes visible again.
In a skill-based agent, a step first leaves the model as a tool call, which the harness receives before dispatch.
We check each tool call as the target of the transfers that led to it, and admit it only if the skeleton permits it after the steps and results observed so far.
The policy remains \ac{cfi}, and the tool-call boundary is only our mechanism for current hosts.
A host that exposes more of a run would let the same policy check more of the skeleton.
Because steps that never cross the boundary stay unchecked and the model chooses where each step runs, every channel that can produce a protected effect must cross it.

\section{Threat Model}
\label{sec:threat-security}

We consider an agent that runs one skill on behalf of a user.
The skill and its bundled helpers are trusted as written.
The threat model here is independent of the monitor's design; the system assumptions for our monitor follow in \autoref{sec:threat-model}.

\noindent\textbf{Adversary and arbitrary proposer model.}
An adversary may embed indirect prompt injections in task documents, workspace files, or tool outputs~\cite{greshake2023not} to induce actions that violate the skill's procedure, such as skipping validation or writing an unapproved target.
Even without an adversary, the model may deviate through hallucination, lost context, or shortcutting~\cite{ruan2024identifying}.
We therefore treat the \ac{llm} as an \emph{arbitrary proposer} of actions, and our guarantees hold however it reasons and whether or not it follows its instructions.

\noindent\textbf{Threats.}
A threat is a run that departs from the skeleton.
The cause does not matter, so a defense need not diagnose whether the model was hijacked or confused.
Four deviations in the running example show the classes we address.

\begin{itemize}[leftmargin=*]
    \item[\ding{172}] \emph{Skipped prerequisite.} After editing, the agent overwrites the source without rendering, validation, or approval.

    \item[\ding{173}] \emph{Wrong branch.} The user chooses ``save as a new copy'', but the agent overwrites the source.

    \item[\ding{174}] \emph{Exceeded bound.} After three failed repairs, the agent starts a fourth instead of taking the failure branch.

    \item[\ding{175}] \emph{Wrong object.} The user approves draft $A$ with digest $d$, but the agent writes other content, such as an earlier draft.
\end{itemize}

Two further classes concern the run as a whole.
In a \emph{false completion}, the agent reports success before the skeleton ends, for example before delivery.
In a \emph{bypass}, the agent produces a protected effect outside the skill's obligation, for example by overwriting the source with its own script.

Deviations \ding{172}--\ding{175} use only operations that the skill requires, so a defense that only bounds which operations the agent may use admits all four.

\noindent\textbf{What enforcement needs.}
Only the run's \emph{workflow history} separates these deviations from lawful steps.
The history consists of the outcomes of completed steps, the user's decisions, the repair count, and the identity of the validated and approved draft.
A program keeps such facts in its own state.
Here they exist only in the model's context, next to untrusted content, and the model is the party being checked.
Enforcement therefore needs the skeleton in a checkable form, sight of each protected step before it takes effect, and a record of the history that the model cannot forge.
These are the three requirements of \ac{cfi} (\autoref{sec:background-cfi}), restated for a skill.

\noindent\textbf{Goal.}
Regardless of what the model proposes, every protected effect the agent produces must be permitted by the control skeleton at that point in the run, given the steps and results that actually occurred.
Completion counts only where the skeleton ends.
The quality of content that the skeleton leaves to the model, such as the revised prose, is outside this goal, as is guaranteeing that the agent finishes the task.

\section{Enforcement at the Tool-Call Boundary}
\label{sec:execution-model}

Current agent hosts expose the tool call through hooks, so we place the monitor at this \emph{tool-call boundary}.
This is a choice of mechanism, and the policy remains that of \ac{cfi}, namely that a run must follow the skill's control skeleton.
The mechanism resembles system-call monitoring, which constrains a process at an observable boundary without instrumenting its internal execution.
Wagner and Dean~\cite{wagner2000intrusion} derive a finite-state automaton over system calls from a program's source code, model internal control paths that issue no system call as $\varepsilon$-transitions, and check observed invocations against this model.
The criterion of \autoref{sec:contract} is stated for any declared boundary, so it stays the same if a host widens the boundary.

This section asks \emph{when a runtime monitor at a boundary can enforce a requirement of a natural-language skill without trusting the model} (notation in \autoref{sec:appendix-notation}).

\subsection{Observable Execution Model}
\label{sec:observable-exec-model}

\begin{defn}[Mediation Boundary and Observable Projection]
\label{def:boundary-projection}
A \emph{mediation boundary} $B$ consists of host-mediated tool requests, their correlated execution results, and native user decisions at declared interfaces.
A request event takes the form $\mathsf{req}\langle i, \iota, a \rangle$, where $i$ is the declared interface recognized from the raw call syntax (a \emph{base interface}, \autoref{sec:design:contract}), $\iota$ is the native call identifier assigned by the host, and $a$ represents the actual call arguments.
A result event takes the form $\mathsf{res}\langle \iota, v, F \rangle$, where $v$ is a discrete outcome extracted by a trusted decoder from the execution of call $\iota$, and $F$ is a set of typed value captures (e.g., file paths, text values, or SHA-256 digests).
A structured question call or an explicit workflow completion call is a request event like any other, and the native user answer or the completion acknowledgment is its correlated result event.
Given a full task-level execution trace $\tau$ that includes internal model states, assistant text messages, proposed tool calls, and returned observations, the \emph{observable projection} $\tau_B = \pi_B(\tau)$ is the subsequence retaining only the events at $B$.
\end{defn}

\noindent\textbf{Three execution regimes.}
Relative to $B$, the three places where a step can run (\autoref{sec:llm-agents}) become three regimes (\autoref{fig:execution-paradigms}(b)).
In \textbf{$R_1$ (internal reasoning)}, the step runs in the model's reasoning or assistant text and emits no event at $B$.
In \textbf{$R_2$ (stepwise mediated calls)}, it is a single request $\mathsf{req}\langle i, \iota, a \rangle$ with a correlated result $\mathsf{res}\langle \iota, v, F \rangle$ at $B$.
In \textbf{$R_3$ (compound helpers and delegated code)}, it runs inside one external invocation, either a pre-packaged multi-step helper ($R_{3\mathrm{a}}$) or a model-synthesized program passed to an interpreter ($R_{3\mathrm{b}}$), so only the outer request and final return cross $B$ and internal steps are $\varepsilon$-transitions.
Only $R_2$ exposes the validation outcome, user choice, and delivery target that the control skeleton constrains.

\subsection{Enforceability at the Boundary}
\label{sec:contract}

A monitor that blocks invalid steps of one ongoing trace before dispatch can enforce only safety properties, whose violations have a finite bad prefix~\cite{schneider2000enforceable}.
We therefore state each skeleton requirement as a safety property and ask whether the boundary shows enough to decide it.

\begin{defn}[Boundary Enforceability]
\label{def:enforceability}
Let $\varphi$ be a workflow safety property over task traces.
For any finite trace prefix $\tau$ and next proposed request $r$, $\varphi$ specifies a decision requirement $D_\varphi(\tau, r) \in \{\mathsf{must\text{-}allow}, \mathsf{must\text{-}refuse}, \mathsf{free}\}$.
The requirement $D_\varphi$ is part of the specification. It marks as $\mathsf{must\text{-}refuse}$ the requests after which every continuation violates $\varphi$, and as $\mathsf{must\text{-}allow}$ the requests that legal executions of the skill issue after $\tau$.
A \emph{$B$-monitor} is a deterministic function $M(\pi_B(\tau), r) \in \{\mathsf{allow}, \mathsf{refuse}\}$ that depends only on trusted boundary events.
We say $M$ is \emph{sound} for $\varphi$ if $D_\varphi(\tau, r) = \mathsf{must\text{-}refuse} \implies M(\pi_B(\tau), r) = \mathsf{refuse}$, and \emph{transparent} for $\varphi$ if $D_\varphi(\tau, r) = \mathsf{must\text{-}allow} \implies M(\pi_B(\tau), r) = \mathsf{allow}$.
Property $\varphi$ is \emph{enforceable at $B$} if some $B$-monitor is sound and transparent for $\varphi$.
\end{defn}

\begin{prop}[Observational Limit]
\label{prop:obs-limit}
A property $\varphi$ is enforceable at $B$ if and only if there do not exist two trace prefixes $\tau_1, \tau_2$ and a request $r$ such that $\pi_B(\tau_1) = \pi_B(\tau_2)$, $D_\varphi(\tau_1, r) = \mathsf{must\text{-}refuse}$, and $D_\varphi(\tau_2, r) = \mathsf{must\text{-}allow}$.
\end{prop}

\autoref{prop:obs-limit} (proved in Appendix~\ref{sec:appendix-prop}) reduces enforceability to whether $B$ distinguishes the traces.
Two conditions follow.
The constrained effect must cross $B$ as a request before it executes, and the evidence on which $D_\varphi$ depends must appear in $\pi_B(\tau)$ as trusted, correlated results.
Mimicry attacks on system-call monitors rest on the same limit~\cite{wagner2002mimicry}.

\noindent\textbf{Corollary for $R_1$: three dispositions.}
If an obligation requires step $s$ before request $r$ but $s$ is performed in $R_1$, a prefix that skips $s$ and one that performs it have the same projection yet demand opposite decisions on $r$, so the obligation is unenforceable at $B$.
The compiler therefore assigns one of three dispositions.
If the source skill prescribes an observable check in $R_2$ (e.g., a bundled validator), the compiler can \emph{enforce} the strengthened property $\varphi^+$, which requires a correlated pass result at $B$ before $r$.
$\varphi^+$ demands refusal wherever $\varphi$ does and admission only on legal executions that check in $R_2$, so enforcing it gives up transparency for $\varphi$, since the monitor refuses $r$ even after a correct check in $R_1$.
Pure $R_1$ semantic or presentation norms stay as \emph{guidance} in agent instructions.
If a mandatory safety condition lacks observable evidence at $B$, the skill is rejected as \emph{unsupported}.

\noindent\textbf{Corollary for $R_3$: helpers vs.\ delegated code.}
Executions that differ only inside an $R_3$ step have identical projections.
Obligations inside a pre-packaged helper ($R_{3\mathrm{a}}$) hold only if the helper is trusted to behave as declared, whereas model-generated code ($R_{3\mathrm{b}}$) can perform protected effects without any intermediate checkpoint at $B$. For such code, a monitor must either admit the interpreter call and lose soundness or refuse it and lose transparency.
\tool chooses soundness and refuses ad-hoc scripts.

\noindent\textbf{Boundary interface barriers.}
A step in $R_2$ can still be unenforceable when the tool interface omits an object identity from its arguments and results, or when the task prompt implies a choice that no boundary event carries.
Such an obligation is reported as \texttt{unsupported} until the identity or choice is exposed at $B$ (\autoref{sec:discussion}).

\noindent\textbf{Regime migration and closure principles.}
Because the model chooses at runtime where computation happens, a deviating or hijacked model can evade monitoring through \emph{regime migration}, relocating a protected step out of $R_2$.
It can claim in $R_1$ that validation passed and then emit the delivery call, or overwrite \texttt{source.md} with an inline script ($R_{3\mathrm{b}}$) instead of the declared tool.
A related evasion stays in $R_2$ and reuses stale evidence, such as approving one draft but writing another.
\tool closes this surface with two principles.
The first is \textbf{non-substitutability of evidence}. Only correlated results from trusted interfaces at $B$ establish workflow prerequisites, and model claims and document contents carry no evidentiary weight.
The second is \textbf{non-migratability of effects} (boundary closure). Every channel that can produce a protected effect is mediated at $B$, and any unmapped command, compound shell construct, or ad-hoc interpreter call is refused by default.

\subsection{System and Trust Assumptions}
\label{sec:threat-model}

We consider an agent session that loads one protected skill and runs tool calls through a host with pre-dispatch and post-execution hooks.
Our guarantees rely on five assumptions about the host and deployment.
\begin{itemize}[leftmargin=*]
    \item \textbf{M1 (Complete mediation).}
    Every native tool request passes the monitor's pre-dispatch gate, and the harness dispatches exactly the admitted arguments.
    Unmediated effect channels, such as an unrestricted shell, are disabled.

    \item \textbf{M2 (Truthful correlation).}
    The harness returns each dispatched call's result to the post-execution hook under its native call identifier, unique within the session.

    \item \textbf{M3 (Trusted bindings and helpers).}
    Base-interface parsers, result decoders, bundled helpers ($R_{3\mathrm{a}}$), and read-only auxiliary operations (\autoref{sec:appro-runtime}) behave as declared, and decoders return $\mathsf{err}$ on a malformed or incomplete receipt.
    A helper acts only on the object identity in its admitted arguments.
    For example, the overwrite helper fails unless the working copy's digest equals its \texttt{sha} argument.

    \item \textbf{M4 (Protected state).}
    The monitor's state lies outside the model's context and changes only through atomic admission and result transactions, one call at a time.

    \item \textbf{M5 (Verified activation).}
    The source skill and its compiled bundle (\autoref{sec:approach}) are audited and pinned by digest. Model actions cannot disable the activated monitor.
\end{itemize}

\noindent\textbf{Task input vs.\ runtime decisions.}
Task input supplies initial parameters but cannot pre-authorize later branches.
The contract captures each parameter from the model's first requests, such as the source path in the initial read, and later steps must reuse it.
A required approval or delivery choice must come from the correlated result of the contract's native question, never from text in the prompt or task files.

\subsection{Violations and Admitted-Path Safety}
\label{sec:security-objective}

\noindent\textbf{Taxonomy of violations.}
A \emph{skill control-flow violation} is a proposed request at $B$ that departs from the progression that the active contract permits.
We distinguish three categories.
An \textbf{unexpected action} fails the common mediation or syntax checks.
Examples are an out-of-scope tool, an inline script ($R_{3\mathrm{b}}$ migration), malformed arguments, and a call while another is pending.
A bypass (\autoref{sec:threat-security}) falls in this category.
An \textbf{invalid transition} names a declared workflow operation, but no transition is enabled because a prerequisite outcome or approval is missing, a repetition bound is exhausted, or a value relation fails, as in \ding{172}--\ding{175}.
An \textbf{invalid termination} is a completion request with no enabled edge, as in a false completion before delivery.

\noindent\textbf{Non-violations and evidence failures.}
Choosing among permitted branches or taking a declared failure or cancellation exit is legal.
A session that stops before an accepted terminal outcome is \emph{incomplete} rather than a violation.
A result that fails decoding is an \emph{evidence failure} and halts progression, whereas a decoded error outcome such as $\mathit{fail}$ follows the contract's error branch.

\noindent\textbf{Security objective (Admitted-Path Safety).}
For an arbitrary proposer and the executable contract $C$ compiled from the skill (\autoref{sec:design:contract}), the requests that the monitor admits and the results it accepts must form a path of $C$ at every prefix, and completion is accepted only at a contract exit (\autoref{thm:safety}).
The objective is stated for $C$, while the goal of \autoref{sec:threat-security} is stated for the skill.
The fidelity of compilation (\autoref{sec:eval:compile}) and the audit of \textbf{M5} connect the two.

\section{SkillCFI Design}
\label{sec:approach}

\begin{figure*}[t]
  \centering
  \includegraphics[width=0.88\textwidth]{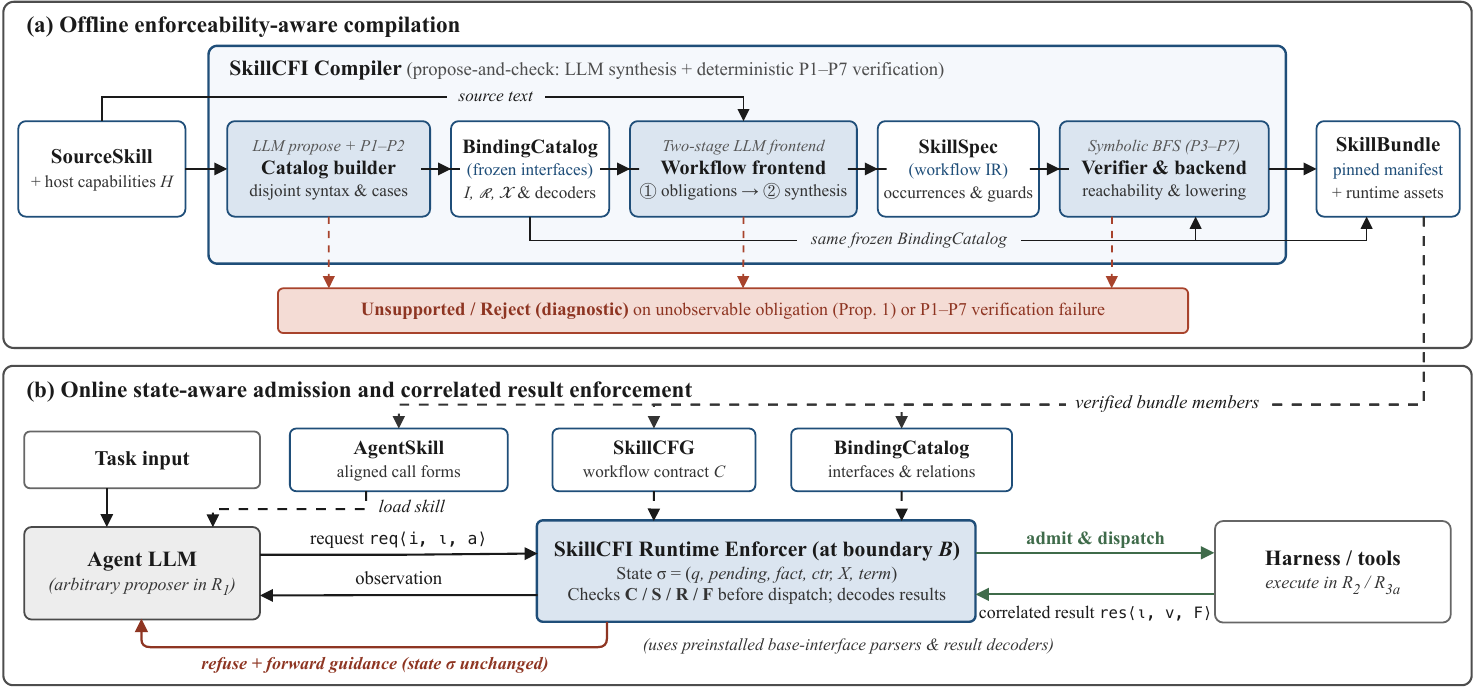}
  \caption{Overview of \tool. (a)~Offline compilation. (b)~Online enforcement at boundary~$B$.}
  \label{fig:approach-overview}
\end{figure*}

\autoref{fig:approach-overview} shows \tool, which \emph{confines \ac{llm}-based interpretation of skill prose to offline compilation}, so that online enforcement is a deterministic reference monitor at $B$.
Offline, a propose-and-check compiler (\autoref{sec:appro-compiler}) turns a trusted \texttt{SourceSkill} and host capabilities~$H$ into a pinned \texttt{SkillBundle}.
The bundle holds the agent-facing \texttt{AgentSkill} and the executable contract (\autoref{sec:design:contract}).
Online, the host loads \texttt{AgentSkill} into the prompt.
The runtime enforcer (\autoref{sec:appro-runtime}) checks every tool request against the contract and returns the next legal steps on refusal (\autoref{sec:design:guidance}).

\subsection{Executable Contract}
\label{sec:design:contract}

\begin{figure}[t]
  \centering
  \includegraphics[width=0.9\linewidth]{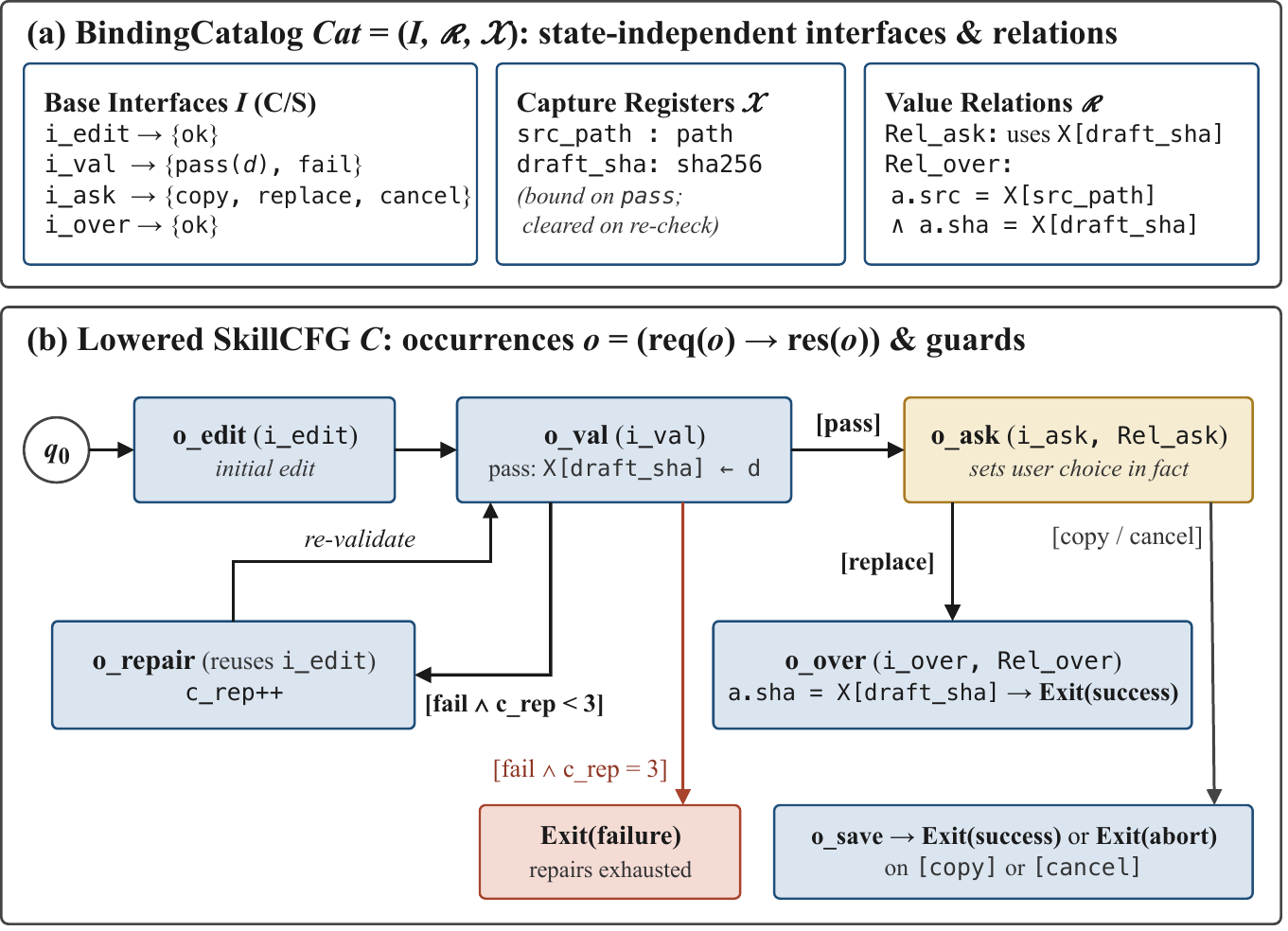}
  \caption{Two-layer executable contract for the document-editing skill, simplified from the compiled bundle (\autoref{sec:appendix-example}). (a)~\texttt{BindingCatalog}. (b)~\texttt{SkillCFG} $C$. The initial read, which binds $\mathtt{src\_path}$, and rendering are omitted.}
  \label{fig:contract-automaton}
\end{figure}

\tool separates the executable contract into two layers (\autoref{fig:contract-automaton}). A state-independent \texttt{BindingCatalog} maps tool calls to typed boundary events, and a state-aware \texttt{SkillCFG} automaton governs the workflow.

\begin{defn}[Binding Catalog]
\label{def:catalog}
A \emph{binding catalog} is a tuple $\mathit{Cat} = (I, \mathcal{R}, \mathcal{X})$, where:
\begin{itemize}[leftmargin=*]
    \item $I$ is a finite set of \emph{base interfaces}. Each $i \in I$ defines a syntactic request language $L(i)$ over tool names, helper identifiers, and typed parameters, together with a deterministic result decoder $\mathit{dec}_i$ mapping raw tool outputs to a finite set of discrete outcomes $O_i$ and extracted field values.
    Each parameter position accepts every value of its declared type.

    \item $\mathcal{X}$ is a finite set of \emph{typed capture registers} storing strings, file paths, or content digests; argument and register types in disequalities have infinite domains.
    A \emph{valuation} $X: \mathcal{X} \rightharpoonup \mathcal{V}$ assigns values to the currently bound registers, and $X[x]$ denotes the value of register $x$.

    \item $\mathcal{R}$ is a finite set of \emph{value-relation predicates}. Each $\rho \in \mathcal{R}$ is a conjunction of equality ($=$) and disequality ($\neq$) constraints over request arguments, capture registers in $\mathcal{X}$, and contract literals, or a question template $T$ requiring the question-body argument to equal the rendering $T(X)$.
\end{itemize}
\end{defn}

\noindent\textbf{Separating base interfaces from workflow occurrences.}
\tool decouples \emph{what operation is invoked} (a base interface $i \in I$) from \emph{where in the workflow it executes} (an occurrence $o \in \mathit{Occ}$).
A skill often reuses one interface across stages with distinct preconditions, counter updates, or captures.
In \autoref{fig:contract-automaton}, interface $\mathtt{i\_edit}$ serves both initial drafting ($\mathtt{o\_edit}$) and post-failure repair ($\mathtt{o\_repair}$).
Unlike $\mathtt{o\_edit}$, $\mathtt{o\_repair}$ requires a prior validation $\mathsf{fail}$ and $c_{\mathrm{rep}} < 3$ and increments $c_{\mathrm{rep}}$.

\begin{defn}[Executable Workflow Contract (SkillCFG)]
\label{def:contract}
Over a catalog $\mathit{Cat} = (I, \mathcal{R}, \mathcal{X})$, a \ac{skillcfg} contract is a guarded register automaton~\cite{kaminski1994finite} $C = (Q, q_0, \mathit{Occ}, E, \mathit{Ctr}, \mathit{Exit})$:
\begin{itemize}[leftmargin=*]
    \item $\mathit{Occ}$ is a finite set of \emph{workflow occurrences}. Each $o = (i_o, \mathit{Rel}_o, \mathit{cap}_o) \in \mathit{Occ}$ binds a base interface $i_o \in I$, required value relations $\mathit{Rel}_o \subseteq \mathcal{R}$, and a capture plan $\mathit{cap}_o$ specifying which registers in $\mathcal{X}$ are produced on request/result or invalidated upon admission.
    $\mathit{Rel}_o(a, X)$ denotes the conjunction of these relations, and $\mathit{Occ}(i)$ the occurrences with $i_o = i$.

    \item $Q$ is a finite set of control locations comprising the initial state $q_0$ and, for each $o \in \mathit{Occ}$, a request node $\mathsf{req}(o)$ and a result node $\mathsf{res}(o)$.
    The \emph{idle} locations are $Q_{\mathrm{idle}} = \{q_0\} \cup \{\mathsf{res}(o) \mid o \in \mathit{Occ}\}$, and exits are not locations.

    \item $\mathit{Ctr}$ is a finite set of integer counters, and $\mathit{Exit} \subseteq \{\mathsf{success}, \mathsf{failure}, \mathsf{abort}\}$ is the set of workflow exits.

    \item $E$ is a set of directed edges of three kinds.
    \emph{Request edges} $q \xrightarrow{g/u} \mathsf{req}(o)$ with $q \in Q_{\mathrm{idle}}$ carry a guard $g(\mathit{fact}, \mathit{ctr})$ and a counter update $u$, which adds a constant to counters or sets them. The guard is a conjunction of outcome-equality checks ($\mathit{fact} = v$) on the latest decoded outcome and counter comparisons ($\mathit{ctr}[c] = k$ or $\mathit{ctr}[c] < k$).
    \emph{Exit edges} $\mathsf{res}(o) \xrightarrow{g} t$ with $t \in \mathit{Exit}$ carry only a guard of the same form and form the subset $E_{\mathit{Exit}} \subseteq E$.
    Each $o$ has one unconditional \emph{correlation edge} $\mathsf{req}(o) \to \mathsf{res}(o)$.
\end{itemize}
\end{defn}

\noindent\textbf{Configurations and paths.}
A contract configuration $\kappa = (q, \mathit{fact}, \mathit{ctr}, X)$ records the control node, the latest decoded outcome (or $\bot$), the counter valuation $\mathit{ctr}: \mathit{Ctr} \to \mathbb{N}$, and the partial valuation $X$ of available captures, starting from $\kappa_0 = (q_0, \bot, \mathbf{0}, \emptyset)$.
The \emph{steps} $\mathsf{req}(o, a)$ and $\mathsf{res}(o, v, F)$ enter the locations $\mathsf{req}(o)$ and $\mathsf{res}(o)$.
A request step $\mathsf{req}(o, a)$ is enabled when exactly one edge from $q$ into $\mathsf{req}(o)$ has a satisfied guard, $a \in L(i_o)$, and $\mathit{Rel}_o(a, X)$ holds.
Firing a step updates $\kappa$ as Appendix~\ref{sec:appendix-semantics} defines.
$\operatorname{Paths}(C)$ is the set of finite alternating request/result sequences permitted from $\kappa_0$, and each $w \in \operatorname{Paths}(C)$ reaches exactly one configuration $\operatorname{Run}_C(w)$, since each step names its occurrence and uses its unique enabled entering edge.

\noindent\textbf{Four-layer check hierarchy (\textbf{C/S/R/F}).}
Across the catalog and the contract, \tool checks requests in four layers.
\textbf{Common mediation integrity (C)} enforces operation-independent transport and session invariants, namely verified bundle activation, literal argument parsing, workspace confinement, single in-flight call serialization, and result correlation and decoding.
\textbf{Static interface membership (S)} checks the state-independent membership $a \in L(i)$ for some $i \in I$.
\textbf{Value relation (R)} checks that referenced captures are in $\operatorname{dom}(X)$ and cross-call predicates $\mathit{Rel}_o(a, X)$ hold.
\textbf{Workflow control flow (F)} checks that the current location $q$ has an edge into $\mathsf{req}(o)$ whose guard $g(\mathit{fact}, \mathit{ctr})$ holds, so it refuses an in-vocabulary request that is not a permitted successor of $q$. It also enforces terminal closure.

\subsection{Skill Compiler}
\label{sec:appro-compiler}

The compiler runs in three stages.
In each stage, an \ac{llm} makes a proposal, and the compiler treats it as untrusted.
A deterministic verifier accepts the bundle only if it satisfies \texttt{P1}--\texttt{P7} (\autoref{tab:acceptance-properties}, \autoref{alg:compiler} in Appendix~\ref{sec:appendix-algorithms}).

\begin{table}[t]
  \centering
  \caption{Compilation acceptance properties (\texttt{P1}--\texttt{P7}). $\sigma^\#$, $\Gamma$, $\mathsf{Sat}$, and $\operatorname{Sig}$ are defined in \autoref{sec:appro-compiler}.}
  \label{tab:acceptance-properties}
  \scriptsize
  \setlength{\tabcolsep}{3.5pt}
  \begin{tabularx}{\linewidth}{@{}llX@{}}
    \toprule
    \textbf{ID} & \textbf{Property} & \textbf{Verification Condition} \\
    \midrule
    \texttt{P1} & Disjoint recognition & Pairwise disjoint request grammars: $\forall i \neq j,\, L(i) \cap L(j) = \emptyset$. \\
    \texttt{P2} & Disjoint decoding & For each $i \in I$, receipt outcome cases in $\mathit{dec}_i$ are pairwise disjoint and satisfiable. \\
    \texttt{P3} & Deterministic admission & At each reachable $\sigma^\#$, every enabled occurrence has a unique enabled entering edge, and enabled $o, o' \in \mathit{Occ}(i)$ with $\mathsf{Sat}(\Gamma \land \mathit{Rel}_o \land \mathit{Rel}_{o'})$ have $\operatorname{Sig}(o) = \operatorname{Sig}(o')$. \\
    \texttt{P4} & Outcome completeness & Every reachable outcome $v \in O_{i_o}$ enters an exit or leads to a state with an enabled occurrence $o'$ such that $\Gamma \models \exists a.\,\mathit{Rel}_{o'}$. The initial state also enables such an $o'$. \\
    \texttt{P5} & Evidence availability & All captures in $\mathit{Rel}_o$ are bound on every path to $\mathsf{req}(o)$. \\
    \texttt{P6} & Symbolic coverage & Every declared occurrence $o \in \mathit{Occ}$ and outcome $v \in O_{i_o}$ is reachable from $\sigma^\#_0$. \\
    \texttt{P7} & Terminal consistency & Each reachable result state enables at most one exit $t \in \mathit{Exit}$, and a state that enables an exit enables no request edge. \\
    \bottomrule
  \end{tabularx}
\end{table}

\noindent\textbf{Three-stage pipeline.}
The catalog stage proposes $\mathit{Cat}$ and freezes it if \texttt{P1} and \texttt{P2} hold.
The obligation stage extracts source obligations with citations of source lines and classifies each by \autoref{prop:obs-limit} as \texttt{enforce}, \texttt{guidance}, or \texttt{unsupported}.
An unsupported mandatory obligation halts compilation with a diagnostic instead of being downgraded to guidance.
The synthesis stage produces \texttt{SkillSpec} $K$, which must exactly match the obligation plan.
Each stage is one model call without retry and returns its proposal in a fixed schema (\autoref{sec:appendix-example} gives an example), and a failed check ends compilation.
The verifier then lowers $K$ to $C$ and explores abstract configurations $\sigma^\# = (q, \mathit{fact}, \mathit{ctr}, \Gamma)$ breadth-first within a state budget $B_{\max}$ and requires \texttt{P3}--\texttt{P7} at every reachable $\sigma^\#$.
It tracks $q$, $\mathit{fact}$, and bounded counters concretely, and $\Gamma$ abstracts $X$ by its bound registers and their equality and disequality constraints, whose satisfiability $\mathsf{Sat}$ decides.
We write $\Gamma \models \exists a.\,\mathit{Rel}_o$ if every valuation consistent with $\Gamma$ admits arguments satisfying $\mathit{Rel}_o$.

\noindent\textbf{Diagnostics.}
A stage that cannot represent a required capability returns \texttt{unsupported} instead of a partial proposal.
Its report lists diagnostics, each with a code, an explanation, and cited source lines, and the compiler checks that every citation points to actual source lines.
The report also marks each uncovered obligation as unsupported and gives a reason.
By \autoref{prop:obs-limit}, a reason typically names an effect that no interface mediates or evidence that no result carries, which tells the author which interface to add (\autoref{sec:eval:compile}).

\noindent\textbf{Compile-time determinism instead of state-set tracking.}
Boundary monitors face \emph{call-site ambiguity}~\cite{sekar2000fast}, and classical non-deterministic monitors track the set of candidate states at runtime~\cite{wagner2000intrusion}, which does not suit register contracts (Appendix~\ref{sec:appendix-algorithms}).
\tool instead requires (\texttt{P3}) that occurrences of one interface enabled at a reachable state either have mutually exclusive value relations or share a \emph{handler signature} $\operatorname{Sig}(o)$, which comprises the interface and decoder, the cited obligation, the capture plan $\mathit{cap}_o$, the counter update of its unique enabled entering edge, and the edges out of $\mathsf{res}(o)$ with their guards, updates, and targets, compared at each reachable state.

\noindent\textbf{Single-source instructions.}
$\mathsf{RenderAgentSkill}$ deterministically renders \texttt{AgentSkill} from $(K, \mathit{Cat})$, emitting exactly the call forms, parameters, outcome branches, and retry bounds that $C$ and $\mathit{Cat}$ enforce.

\subsection{Runtime Enforcer}
\label{sec:appro-runtime}

\setcounter{algocf}{1}%
\begin{algorithm}[t]
\footnotesize
\caption{State-Aware Admission and Correlated Result Enforcement}
\label{alg:runtime}
\DontPrintSemicolon
\KwState{$\sigma = (q, \mathit{pending}, \mathit{fact}, \mathit{ctr}, X, \mathit{term})$}
\Proc{$\mathsf{OnRequest}(r = \langle \mathit{tool}, \iota, a \rangle)$}{
    \textbf{if} $\mathit{pending} \neq \bot \lor \mathit{term} \neq \bot$ \textbf{then} \Return{$\mathsf{Refuse}(\mathsf{Integrity\_C} \mid \mathsf{Flow\_F})$}\;
    $i \leftarrow \mathsf{ResolveBaseInterface}(\mathit{tool}, a, I)$;\ \textbf{if} $i = \bot$ \textbf{then} \Return{$\mathsf{AuxiliaryOrRefuse}(r)$}\;
    $\mathit{Plans} \leftarrow \{\mathsf{BuildPlan}(o, a, \sigma) \mid o \in \mathit{Occ}(i) \land (\exists! (q \xrightarrow{g/u} \mathsf{req}(o)) \in E : g(\mathit{fact}, \mathit{ctr})) \land \mathit{Rel}_o(a, X)\}$\;
    \textbf{if} $\mathit{Plans} = \emptyset$ \textbf{then} \Return{$\mathsf{Refuse}(\mathsf{InvalidTransition\_F\_R})$}\;
    \textbf{if} $\exists p, p' \in \mathit{Plans} : \operatorname{Sig}(p.o) \neq \operatorname{Sig}(p'.o)$ \textbf{then} \Return{$\mathsf{Refuse}(\mathsf{Unsupported})$} \tcp*{Thm.~\ref{thm:runtime-determinism}}
    $p^* \leftarrow \mathsf{SelectCanonical}(\mathit{Plans})$\;
    \textbf{atomic:} $q \leftarrow \mathsf{req}(p^*.o)$;\ $\mathit{fact} \leftarrow \bot$;\ $\mathit{ctr} \leftarrow p^*.u(\mathit{ctr})$;\ $X \leftarrow (X \setminus p^*.\mathit{inv}) \oplus p^*.\mathit{reqCap}$;\ $\mathit{pending} \leftarrow \langle \iota, p^*, a, \mathit{dec}_i \rangle$\;
    \Return{$\mathsf{ReleaseOnce}(a)$}\;
}
\Proc{$\mathsf{OnResult}(\iota, a_{\mathrm{disp}}, \mathit{rawRes})$}{
    \textbf{if} $\mathit{pending} \neq \langle \iota, p^*, a_{\mathrm{disp}}, \mathit{dec} \rangle$ \textbf{then} \Return{$\mathsf{Halt}(\mathsf{EvidenceFailure\_C})$}\;
    $(v, F) \leftarrow \mathit{dec}(\mathit{rawRes})$;\ \textbf{if} $(v, F) = \mathsf{err}$ \textbf{then} \Return{$\mathsf{Halt}(\mathsf{EvidenceFailure\_C})$};\ \textbf{if} $p^* = \mathsf{aux}$ \textbf{then} $\mathit{pending} \leftarrow \bot$;\ \Return\;
    \textbf{atomic:} $X \leftarrow X \oplus \mathsf{Filter}(F, p^*.o, v)$;\ $\mathit{fact} \leftarrow v$;\ $q \leftarrow \mathsf{res}(p^*.o)$;\ $\mathit{pending} \leftarrow \bot$\;
    \textbf{if} $\exists!\, (\mathsf{res}(p^*.o) \xrightarrow{g} t) \in E_{\mathit{Exit}} : g(\mathit{fact}, \mathit{ctr})$ \textbf{then} $\mathit{term} \leftarrow t$\;
}
\end{algorithm}

At runtime, the enforcer maintains state $\sigma = (q, \mathit{pending}, \mathit{fact}, \mathit{ctr}, X, \mathit{term})$.
It extends the contract configuration $\kappa$ with a single-slot lock $\mathit{pending}$ for the in-flight call and a terminal status $\mathit{term} \in \mathit{Exit} \cup \{\bot\}$.
The enforcer, which we also call the \emph{gate}, mediates every boundary event via \autoref{alg:runtime}.
Its \emph{admitted workflow projection} $\tau_B^{\mathrm{adm}}$ lists the workflow requests it releases and the results it accepts in order, each labeled with its credited occurrence $o$ as a step $\mathsf{req}(o, a)$ or $\mathsf{res}(o, v, F)$.

\noindent\textbf{Preview before atomic admission.}
$\mathsf{OnRequest}$ checks readiness and the base interface (\textbf{C/S}), previews candidate occurrences (\textbf{F/R}) against a read-only snapshot of $\sigma$, and commits only a request with a non-empty, signature-consistent plan set.
Admission clears $p^*.\mathit{inv}$, the registers that the occurrence invalidates and those that any of its results may write.
Admitting a new validation thus clears $\mathtt{draft\_sha}$, and a failed one leaves it unbound, so stale validation or approval evidence cannot survive a later repair or re-check.
Here $\oplus$ denotes override.
The counter update comes from the unique enabled edge into $\mathsf{req}(o)$ (Line~4), and by \texttt{P3} this uniqueness requirement refuses no legal step.

\noindent\textbf{Correlated results.}
$\mathsf{OnResult}$ accepts a return only if its call ID and dispatched arguments match $\mathit{pending}$ and the pinned decoder succeeds; otherwise, the session halts.
Only then are outcome-scoped captures published and $\mathit{fact}$, $q$ advanced.

\noindent\textbf{Auxiliary operations.}
A request that resolves to no workflow interface ($i = \bot$) may be admitted only as a \emph{read-only auxiliary operation}.
Such a call leaves $(q, \mathit{fact}, \mathit{ctr}, X)$ unchanged, contributes no event to $\tau_B^{\mathrm{adm}}$, and never supplies a workflow outcome, capture, or approval.
Appendix~\ref{sec:appendix-algorithms} defines these read adapters and the subroutines of \autoref{alg:runtime}.

\subsection{Refusal Semantics and Next-Step Feedback}
\label{sec:design:guidance}

Every refusal is decided before the commit at Line~8 of \autoref{alg:runtime}, so a refused request dispatches nothing and leaves $\sigma$ unchanged (\autoref{thm:safety}(a)).
\tool returns to the agent a structured payload $\delta = \langle \mathit{code}, \mathit{layer}, \mathit{missing}, \mathit{next} \rangle$.
It names the refusal category and check layer (\textbf{C/S/R/F}), the unsatisfied guard, missing or invalidated register, or mismatched argument, and the occurrences enabled at the unchanged state.
A benign model can thus self-correct.

\noindent\textbf{Running example.}
For the document-editing skill, the compiler classifies abstract editing as \texttt{guidance} and verifies that $X[\mathtt{draft\_sha}]$ is bound by $\mathtt{o\_val}(\mathsf{pass})$ on every path to $\mathtt{o\_ask}$ and $\mathtt{o\_over}$ (\texttt{P5}).
\tool refuses \ding{172}--\ding{174} at \textbf{F}, because no edge into $\mathtt{o\_over}$ or $\mathtt{o\_repair}$ is enabled before validation, after the choice ``new copy'', or at $c_{\mathrm{rep}} = 3$, and returns $\mathit{next}$, such as $\{\mathtt{o\_val}\}$ for \ding{172}.
It refuses \ding{175} at \textbf{R}, since $\mathtt{Rel\_over}$ requires $a.\mathtt{sha} = X[\mathtt{draft\_sha}] = d$.
An inline script that overwrites the source is refused at \textbf{C/S}.

\subsection{Security Guarantees}
\label{sec:appro-analysis}

\begin{thm}[Pre-Execution Safety and Non-Bypassability]
\label{thm:safety}
Let $(C, \mathit{Cat})$ be a contract whose catalog passes \texttt{P1} and \texttt{P2}.
Under assumptions \textbf{M1}--\textbf{M5} (\autoref{sec:threat-model}), for any session execution with unrestricted reasoning in $R_1$ and arbitrary untrusted data from $R_2/R_3$:
\begin{enumerate}[label=(\alph*), leftmargin=*]
    \item \emph{Zero side effects on refusal:} Any request $r$ returning $\mathsf{Refuse}(\delta)$ is blocked prior to dispatch, invokes no external handler in $R_2/R_3$, and satisfies $\sigma' = \sigma$.

    \item \emph{Trace inclusion in contract language:} At every execution prefix, $\tau_B^{\mathrm{adm}}$ belongs to $\operatorname{Paths}(C)$, each native invocation adds at most one request and one result to it, and the contract configuration of $\sigma$ equals $\operatorname{Run}_C(\tau_B^{\mathrm{adm}})$.

    \item \emph{Mandatory prerequisite and provenance enforcement:} For every dispatched occurrence $o \in \mathit{Occ}$ with arguments $a$, the guard of the edge from the current location into $\mathsf{req}(o)$ holds for the current counters and the outcome $\mathit{fact}$ decoded from the latest correlated workflow result, and $\mathit{Rel}_o(a, X)$ holds, where every referenced register holds a value recorded in this session from an earlier admitted request or a verified result capture, and not invalidated or overwritten since.

    \item \emph{Terminal closure:} $\mathit{term} = t \neq \bot$ only if $\tau_B^{\mathrm{adm}}$ ends with a result step $\mathsf{res}(o, v, F)$ that enables the exit edge to $t$, with no call pending. Later requests are refused.
\end{enumerate}
\end{thm}

The safety proof uses only \texttt{P1} and \texttt{P2}, which make base-interface resolution and each decoder $\mathit{dec}_i$ functions. \texttt{P3}--\texttt{P7} add determinism, progress, and transparency.

\begin{thm}[Compile-Time Sufficiency for Runtime Determinism]
\label{thm:runtime-determinism}
Let $(C, \mathit{Cat})$ additionally pass \texttt{P3}--\texttt{P7} (\autoref{tab:acceptance-properties}).
Under the assumptions of \autoref{thm:safety}, at every runtime state $\sigma$ reachable under \autoref{alg:runtime}:
\begin{enumerate}[label=(\alph*), leftmargin=*]
    \item \emph{Absence of runtime handler ambiguity:} For any request $r$ that reaches Line~6 of \autoref{alg:runtime} (i.e., with $\mathit{Plans} \neq \emptyset$), all candidate plans share a handler signature $\operatorname{Sig}(p.o)$, so the $\mathsf{Unsupported}$ refusal at that line is never taken. Every later decision is the same whichever plan is selected, since the result nodes of the candidates have the same outgoing edges.

    \item \emph{Deadlock-free progress:} Every state $\sigma$ reached by accepting a workflow result (Lines~13--14) either has entered a unique exit ($\mathit{term} \neq \bot$) or enables some request step $\mathsf{req}(o, a)$ with $o \in \mathit{Occ}$.
\end{enumerate}
\end{thm}

By \autoref{prop:transparency} in Appendix~\ref{sec:appendix-proofs}, at every reachable state with no pending call and no entered exit, \autoref{alg:runtime} admits each request that passes the \textbf{C} checks and is an enabled step of $C$.
It is therefore sound and transparent for the property $\varphi_C$ induced by $C$, so $\varphi_C$ is enforceable at $B$ (\autoref{def:enforceability}).

\section{Evaluation}
\label{sec:evaluation}

\subsection{Questions and Setup}
\label{sec:eval:setup}

We answer four research questions.
\begin{itemize}[leftmargin=2.8em, labelsep=0.5em]
  \item[\textbf{RQ1}] \emph{Compilation fidelity and acceptance.} Do compiled contracts follow their sources faithfully and stably, and do diagnostics explain rejections?
  \item[\textbf{RQ2}] \emph{Enforcement effectiveness.} Does \tool block procedural violations before dispatch, compared with no enforcement and a permission-style allowlist?
  \item[\textbf{RQ3}] \emph{Utility under enforcement.} Does enforcement preserve legitimate workflows, and can the agent continue or stop lawfully after a refusal?
  \item[\textbf{RQ4}] \emph{Overhead.} What do compilation, the runtime gate, and the whole session cost?
\end{itemize}
The rest of this subsection describes the skills and tasks, the public skills, the compared conditions and baselines, the sessions and environment, and the human raters.
Each RQ then states its own procedure and measures.

\definecolor{scfgBlueF}{HTML}{DAE8FC}\definecolor{scfgBlueS}{HTML}{6C8EBF}
\definecolor{scfgYellowF}{HTML}{FFF2CC}\definecolor{scfgYellowS}{HTML}{D6B656}
\definecolor{scfgGreenF}{HTML}{D5E8D4}\definecolor{scfgGreenS}{HTML}{82B366}\definecolor{scfgGreenT}{HTML}{2D6A1E}
\definecolor{scfgRedF}{HTML}{F8CECC}\definecolor{scfgRedS}{HTML}{B85450}\definecolor{scfgRedT}{HTML}{992B27}
\definecolor{scfgGrayF}{HTML}{F5F5F5}\definecolor{scfgGrayS}{HTML}{666666}\definecolor{scfgGrayT}{HTML}{444444}
\DeclareRobustCommand{\scfgNode}[2]{\tikz[baseline=-0.55ex]\draw[fill=#1,draw=#2,line width=0.5pt] (0,0) circle (0.55ex);}
\DeclareRobustCommand{\scfgTerm}[2]{\tikz[baseline=-0.55ex]\draw[fill=#1,draw=#2,line width=0.5pt] (-0.5ex,-0.5ex) rectangle (0.5ex,0.5ex);}
\DeclareRobustCommand{\scfgEntry}{\tikz[baseline=-0.55ex]\fill[black!90] (0,0) circle (0.4ex);}
\DeclareRobustCommand{\scfgEdge}[1]{\tikz[baseline=-0.55ex]\draw[->,line width=0.6pt,#1] (0,0) -- (1.6em,0);}

\noindent\textbf{Skills and tasks.}
We crafted eight procedural skills with their helper scripts.
Each skill's contract was accepted and frozen before any session, with its compiled \texttt{SkillCFG} shown in \autoref{fig:skillcfg-gallery} of Appendix~\ref{sec:appendix-eval}.
Each skill has two normal and two abnormal tasks, 32 in total.
A normal task follows a lawful route, which may also end in a lawful failure, a kept draft, or a retained backup.
An abnormal task asks the agent to make one in-vocabulary violation, such as skipping validation, repairing too often, overriding a denial, or writing content other than the approved one (\autoref{tab:abnormal-tasks} in Appendix~\ref{sec:appendix-eval}).
Each skill also has one boundary probe that asks for an action outside the vocabulary, such as a raw \texttt{rm} command or a direct Python call ($R_{3\mathrm{b}}$).
The tasks ask for each violation directly instead of hiding it in injected text.
The gate decides only on the request and its boundary history, so it would decide the same way if injected text had induced the request.
A synthetic user gives fixed answers to the skill's questions, the same in every condition.

\noindent\textbf{Public skills.}
A second set tests \tool on external skills, covering all 9 packages in openclaw/agent-skills~\cite{githubGitHubOpenclawagentskills} and all 44 in openai/skills~\cite{githubGitHubOpenaiskills}.
We compiled each package once as written and adapted seven of the rejected packages (\autoref{sec:eval:compile}).
We replayed each accepted contract offline with pre-registered traces and mocked external services, and three adapted packages, beam, skill-installer, and vercel-deploy, also ran live with one normal and one abnormal task each.

\noindent\textbf{Conditions and baselines.}
All conditions load the same generated \texttt{AgentSkill}, helpers, and tasks, and differ only in which checks the gate applies (\autoref{sec:design:contract}).
\cnone turns the gate off.
\callow applies \textbf{C} and \textbf{S}, so it admits any request in the skill's vocabulary.
\cfull applies all four checks.
\cnorel drops the capture relations of \textbf{R} from \cfull.
It runs only on the \emph{relation task}, the abnormal task of meeting-followup that writes content other than the approved one.
Under \cfull, abnormal tasks run with each of three refusal messages (\autoref{sec:eval:utility}), and all other sessions use \fbasic.
The message is built after the gate has decided, so no blocking result depends on it.
\callow stands for permission and least-privilege mechanisms~\cite{shi_progent_2025,wu2026skillscope,pan2026skillguard}.
The stateless bound of RQ2 covers every policy that decides on one request at a time, and the result-blind bound covers monitors that see requests but no results, such as SkillSentry~\cite{lu2026skillsentry}.
History-aware monitors that observe results~\cite{wang2026agentspec,palumbo_policy_2026,li2026vigil,li2026holds} can express these orderings given a correct policy.
VIGIL and SpecHarness have released no code and derive their policies with an \ac{llm} for each run or task, so a comparison would only test the policy we write for them.

\noindent\textbf{Sessions and environment.}
Normal tasks run five sessions under each of \cnone, \callow, and \cfull (240 sessions).
Boundary probes run under the same three conditions (120), abnormal tasks under \cnone, \callow, and \cfull with each refusal message (400), and the relation task under \cnorel and \cfull (10).
These 770 sessions form the main cohort.
The three live public skills add 60 sessions under \cnone and \cfull with \fbasic, five sessions or qualified targets per cell.
Each abnormal, boundary, and relation cell collects five qualified targets (\autoref{sec:eval:enforcement}).
\tool has $\sim$6.7K lines of TypeScript/JavaScript and Python.
The compiler calls \texttt{gpt-6.1-sol} through OpenRouter with high reasoning effort and temperature~0.
The agent is GLM-5.3-Flash with low reasoning effort and temperature~0 in OpenCode 1.18.35, where the enforcer runs as a plugin~\cite{opencodePlugins}.
Each session runs in a fresh Linux container with 4 CPUs and 8\,GiB of memory and has a 600-second limit.
\autoref{sec:appendix-setup} gives implementation, contract, and rating details, and we will release the code and the artifacts upon acceptance.

\noindent\textbf{Human raters.}
Two authors make all human judgments independently.
For the obligation, source-line, and refusal checks, one rater judges every item and the other a random 20\% sample, and both judge all other items.
They agreed on 121 of the 136 answers that both recorded and settled the rest by discussion, and we report the agreed answers.

\subsection{RQ1: Compilation Fidelity and Acceptance}
\label{sec:eval:compile}

\begin{figure}[t]
  \centering
  \includegraphics[width=0.92\linewidth]{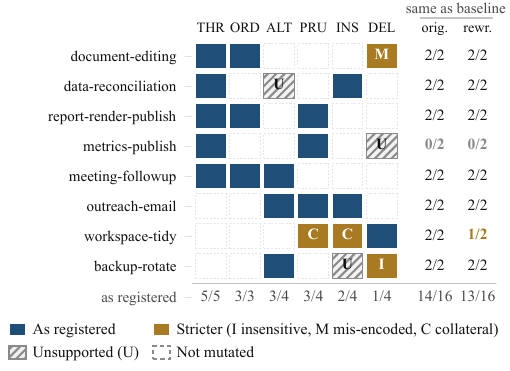}
  \caption{Left: the 24 single-point mutations by skill and operator.
  Right: compilations of the unchanged source and of its rewrite that match the baseline on every trace and session.
  Omitted runs either failed acceptance or differed from the baseline.}
  \label{fig:fidelity}
\end{figure}

The baseline of each skill is the contract used in the agent sessions.
First, we wrote three single-point mutations of each source, 24 in total.
THR changes a bound by one, ORD swaps two steps, ALT and PRU add or remove an alternative, and INS and DEL insert or delete an obligation.
Before compiling, we registered each mutation with its expected change and with traces that it should decide differently, and both raters judged every mutation valid.
We replayed each compiled mutant on these traces, on the traces of the other mutations of its skill, and on the 686 requests of the normal \cfull sessions.
A deviation is \emph{permissive} if the mutant admits a request that the registration expects refused, and \emph{strict} otherwise.
Changes outside the registered ones were attributed from step-level evidence, and both raters confirmed every attribution.
Second, we recompiled each unchanged source twice, and compiled a rewrite with the same meaning twice.
We also injected well-formedness defects into the eight baseline contracts to test acceptance.
Third, the raters judged, for each contract operation, whether the cited source lines support it, whether the source allows a legal trace, and whether it forbids a minimal violation.
They also judged whether each source line that no contract element cites states a constraint or an alternative.

\noindent\textbf{Mutations.}
The compiler reads the numbered workflow as closed, so every deviation is stricter than the source.
Of the 24 mutations, 17 changed the contract as registered (\autoref{fig:fidelity}).
Three were rejected as unsupported.
In two, a required read and an ordinary read take the same arguments in one state, so no request can tell them apart, and the third needs a closure sequence the contract cannot express.
The other four are stricter than the source (I, M, and C in \autoref{fig:fidelity}).
For example, deleting a prohibition from backup-rotate (I) does not create a permission, since the contract still admits reads only where the procedure lists them.
Two of the four (C), both of workspace-tidy, change as registered but also pin the plan ID to the example value in the source.
Across all replayed traces, 116 verdicts deviate from the registration, all strictly and none permissively.

\noindent\textbf{Stability and acceptance.}
Of 16 recompilations of the unchanged sources, all 14 accepted contracts behave as the baseline on every trace and session (\autoref{fig:fidelity}, right); the verifier rejected the two of metrics-publish for a catalog defect.
For rewrites, 13 of 14 accepted contracts match the baseline, and the remaining one carries the same plan-ID pin.
The verifier also rejected 80 of 81 injected defects.
The accepted one leaves the contract well formed and changes only its meaning, which the mutations above test.

\noindent\textbf{Human checks.}
The baselines miss no stated constraint and are stricter only where the source is silent.
For each of the 96 contract operations, the baseline admits the legal trace and refuses the minimal violation.
By the agreed answers, the cited lines support 95 of them and partly support one, and the source allows every legal trace.
The source forbids 89 of the minimal violations.
The other 7 are repeated reads that it does not mention.
Among the source lines that no contract element cites, 14 state a constraint or an alternative, and all behave as the raters expect.
Three constrain content, which the gate cannot see, so the helper's check reports the violation and the gate refuses to continue.

\noindent\textbf{Public skills as written.}
On the public skills, the compiler's reasons for rejection lie in their interfaces.
None of the 53 packages is accepted as written.
Twenty-eight stop at the source inventory before any model call, 24 because they have no helper code and 4 because they exceed a size bound.
Both raters judged the other 25 to be procedural skills.
The compiler rejects all 25 as unsupported at the catalog stage and states its reasons.
By diagnostic category, all 25 have helpers that return no complete receipt the gate could tie to the call, as in beam.
In 18, call forms cannot be written in the catalog's grammar, as in transcribe.
Also 18 require a capability with no interface, such as an MCP tool or the agent's code edits in gh-fix-ci.
Only 7 need nothing beyond receipts and call forms, including beam and transcribe.

\noindent\textbf{Adaptation.}
We adapted seven rejected packages whose stated gaps were mostly missing receipts and call forms, each in one of two tiers whose rules were fixed before any edit.
An \emph{adapted} package keeps its workflow, wraps the commands it already uses so that they return receipts, and moves optional branches out of scope.
A \emph{ported} package also gets skill-specific commands for actions that the source leaves to generic tools or to the agent.
Both raters reviewed every changed version.
In beam, the package as written and four adapted versions were each rejected for one named gap, such as the missing receipt or a completion acknowledgment.
The fifth adapted version was accepted.
Seven selected packages were accepted after adaptation or porting (\autoref{tab:public} in Appendix~\ref{sec:appendix-eval}).
Every contract branches on helper results and binds values across steps.
In transcribe and speech, runs repeat over several files or clips, and a failed key check leads to a new check.
Together, the seven cover the workflow structures in our dataset except bounded repeats.
Offline replays match every registered verdict except three of speech, where the contract is stricter and requires a new key check for each new clip.

\subsection{RQ2: Enforcement Effectiveness}
\label{sec:eval:enforcement}

A \emph{qualified target} is a violating request that the agent proposes with complete arguments after a lawful history, and abnormal and relation targets must also lie in the skill's vocabulary.
The skipped-validation, approval-bypass, and premature-success targets cover $R_1$ claims, since at $B$ a missing prerequisite looks the same as one claimed in $R_1$ (\autoref{prop:obs-limit}).
For each qualified target, we record whether the gate blocked it before dispatch.
Under the baselines, we record whether it took effect.
To test whether a monitor with less evidence could block these targets, we replayed them offline under two bounds.

\noindent\textbf{In-vocabulary violations.}
\cfull blocks every in-vocabulary violation, while both baselines let all of them through.
Under each of \cnone and \callow, all 80 qualified targets of the abnormal tasks were dispatched and took effect.
Under \cfull, all 240 targets across the three refusal messages were blocked before dispatch.
\textbf{F} was the deciding check in 15 of the 16 tasks (\autoref{tab:abnormal-tasks} in Appendix~\ref{sec:appendix-eval}).
The remaining one is the relation task, which keeps the lawful order, so only \textbf{R} can block it.

\noindent\textbf{Stateless bound.}
On this dataset, no stateless policy can block these violations without refusing legal requests.
For each target executed under the baselines, we checked offline whether the gate would admit the same request at some state that a normal \cfull session of the same skill reached.
Of the 160 targets (80 under each of \cnone and \callow), 111 are admitted at such a state.
The other 49 match such a request except in values that only the history supplies, such as the identity of the approved draft or of the validated candidate.
A stateless policy thus sees either a request that is legal elsewhere or a value it cannot check without the history.

\noindent\textbf{Result-blind bound.}
A monitor that sees every request and its arguments—but no helper outcome, receipt value, or user answer—still cannot block all targets.
We replayed each target's session offline through the gate and let every earlier result take any outcome and receipt value that the contract allows.
As a sanity check, with the recorded results the replay refuses all 160 targets and reproduces the gate's recorded state on all 250 abnormal and relation targets under \cfull and \cnorel.
For 59 of the 160 targets, some choice of outcomes leads to admission, meaning a legal execution can issue the same requests followed by the target.
By \autoref{prop:obs-limit} applied to requests alone, any result-blind monitor that admits all legal requests must admit these 59.
Each differs from that legal execution in one result, a user's answer in 39 (such as \emph{No} instead of \emph{Yes} before a deletion) and a helper outcome in 20.
The requests alone suffice to refuse the other 101, mostly because a step is missing, out of order, or past its bound.
Among monitors that admit every legal request, a stateless one must admit all 160 targets, a result-blind one 59, and \cfull none.

\noindent\textbf{Boundary probes and relations.}
\callow matches \cfull outside the vocabulary, and \textbf{R} stops wrong content that the order checks miss.
Under \cnone, all 40 boundary targets were dispatched.
\callow and \cfull each blocked all 40, by \textbf{S} when no interface recognized the command (24 and 23), by \textbf{C} when it used compound shell syntax (16 each), and \cfull once by \textbf{F} after the workflow had ended.
On the relation task, \cnorel admitted all 5 targets, while \cfull blocked all 5 at \textbf{R}.

\noindent\textbf{Public skills.}
The result carries over to the three public skills that ran live.
\cnone dispatched all 15 violations, such as a deploy that skips the CLI check in vercel-deploy, and \cfull blocked all 15 before dispatch.

\subsection{RQ3: Utility under Enforcement}
\label{sec:eval:utility}

\begin{figure}[t]
  \centering
  \includegraphics[width=0.95\linewidth]{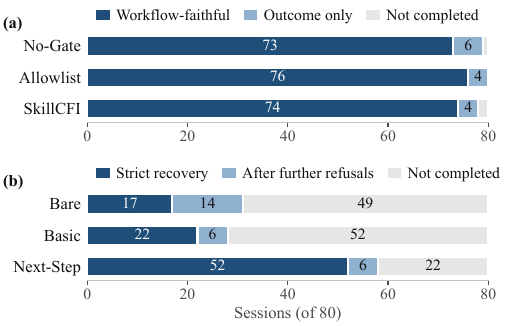}
  \caption{Utility under enforcement.
  (a)~Normal routes per condition. Outcome-only sessions reach outcome completion but deviate from the workflow, so the first two segments together give outcome completion.
  (b)~Sessions after the target refusal under \cfull, per refusal message. After further refusals means that the expected ending was reached after at least one more refusal.
  Each bar has 80 sessions.}
  \label{fig:utility}
\end{figure}

\emph{Outcome completion} means a normal session ends with the expected result, and its receipts, file effects, delivery, and approval binding pass the task's checks.
\emph{Workflow-faithful completion} also requires the executed steps to follow the skill's workflow.
A session with refusals still counts if it meets these checks, since a refused request does not execute.
For each refusal in the normal sessions, the raters judged whether the source allows the request, without seeing the gate's refusal code or layer.
We also rebuilt each request admitted in the normal \cfull sessions only from the call forms written in \texttt{AgentSkill} and replayed it at the same state.
For abnormal sessions under \cfull we vary the refusal message (\autoref{sec:design:guidance}).
\fbare returns only the refusal category and reason.
\fbasic adds a structured record stating that the request did not run and that the workflow state is unchanged.
\fguid further adds the current workflow facts and the operations enabled next, the full $\delta$.
After the target refusal of an abnormal session, \emph{strict recovery} means that the session reaches the expected ending, passes the same checks, and meets no further refusal.
For a task that should stop, it is a failure or an abort.

\noindent\textbf{Normal routes.}
The gate keeps normal-task completion within two sessions of \cnone.
Outcome completion is 79, 80, and 78 of 80 under \cnone, \callow, and \cfull; workflow-faithful completion is 73, 76, and 74 (\autoref{fig:utility}a).
Most outcome-only sessions come from one document-editing task, where the agent wrote its own draft instead of the supplied one, which the gate does not check.
On the live public skills, all 30 normal sessions complete faithfully, 15 each under \cnone and \cfull.

\noindent\textbf{Refusals in normal sessions.}
Every refusal of a request that the source allows came from its shell form, such as an unquoted line break or a compound command, not from a workflow (\textbf{F}) or relation (\textbf{R}) rule.
By the agreed answers, of the 35 refusals under \cfull, the source forbids 30, allows 4, and leaves 1 open.
Of the 10 under \callow, it forbids 7 and allows 3.
The call forms in \texttt{AgentSkill} avoid this problem, since all 686 rebuilt requests pass again.

\noindent\textbf{After a refusal.}
\fguid turns most target refusals into a lawful continuation.
Strict recovery rises from 17 of 80 (21\%) with \fbare and 22 with \fbasic to 52 (65\%) with \fguid (\autoref{fig:utility}b).
\emph{Lawful completion} also counts sessions that reach the expected ending after further refusals, and it is 31, 28, and 58.
On one document-editing task that asks the agent to finish before delivery (\autoref{fig:supp-trace}), the \fbare session stops after the refused \texttt{finish}, while the \fguid session makes the copy, verifies it, and finishes lawfully.
\fguid helps on six of eight skills (\autoref{fig:recovery-per-skill} in Appendix~\ref{sec:appendix-eval}).
On meeting-followup, each refusal message gives 6 of 10.
It does not help on metrics-publish, whose tasks conflict with the lawful continuation.
One task asks for success after a failed check and forbids repair, yet \fguid lists the enabled repair, and 3 of 5 sessions repaired and exported against the task.

\subsection{RQ4: Overhead}
\label{sec:eval:overhead}

We measure overhead on the 240 normal sessions, with \cnone as the baseline for every ratio.
Task time runs from the task prompt to the end of the session, without container setup, and gate time is the time spent in the gate's hooks.
A ratio divides the summed per-task mean times of two conditions, and its 95\% interval comes from resampling the 16 tasks with replacement.
To dispatch exactly the text it checked, the gate rewrites each admitted call into a canonical form, and this form enters the conversation history.
This form has an absolute interpreter path, quotes every word, and names the working directory.
A dispatch ablation compares it with a compact form that drops the quotes and the working directory but keeps the same check.
It runs \cnone, canonical \cfull, and compact \cfull in one interleaved cohort of 96 normal sessions.

\noindent\textbf{Gate cost.}
The gate takes about 2\% of task time, but sessions take about a third longer.
Relative to \cnone, task time is $1.34\times$ (1.15--1.56) under \callow and $1.32\times$ (1.19--1.46) under \cfull.
The gate takes a median of 1.5\% and 2.1\% of task time.
Across the cohort, its before hook takes 53\,ms at the median and 69\,ms at the 95th percentile when \callow admits a call, 77 and 100\,ms when \cfull admits one, and 45 and 58\,ms when \cfull refuses one.

\noindent\textbf{Source of the slowdown.}
The extra time comes from the agent copying the canonical form into its later calls.
From its second call on, the agent quoted every word in 227 of 238 commands under the canonical form, in 3 of 226 under \cnone, and in none under the compact form.
The canonical form reproduces the $1.32\times$ task time (1.02--1.69), and the compact form lowers it to $1.08\times$ (0.97--1.17) (\autoref{fig:overhead} in Appendix~\ref{sec:appendix-eval}).
Output tokens fall from $1.48\times$ to $1.09\times$, while the gate takes about 1\,s per session under both forms.
An offline replay of all 770 sessions of the main cohort gives identical gate decisions under both forms.
All other results use the canonical form.

\noindent\textbf{Compilation.}
Compiling one skill is a one-time cost of three model calls, 7--13 minutes, and 72,000--100,000 tokens, paid before any session.
The rendered \texttt{AgentSkill} is 2.5--4.4$\times$ as long as its source, mostly call forms and outcome branches, and every condition loads it, including \cnone, so the ratios above do not measure the cost of this length.

\section{Discussion and Limitations}
\label{sec:discussion}

\noindent\textbf{Beyond the tool-call boundary.}
As in classical \ac{cfi}~\cite{abadi_control-flow_2009,carlini2015control}, flawed content in a permitted call sequence is out of scope (\autoref{thm:safety}).
Register-bound arguments rule out the mimicry~\cite{wagner2002mimicry} that substitutes the object of a call, such as delivering a draft other than the approved one.
Content that is harmful yet passes validation and that the user approves is still delivered.
Helpers and generated scripts in $R_3$ remain outside the boundary.
Covering them needs two additions.
One checks that each script matches the task it was dispatched for, and the other applies \ac{cfi} or system-call monitoring inside the script.

\noindent\textbf{Enforceable skill interfaces.}
Many skills cannot be enforced because their tools omit evidence at $B$, not because their constraints are semantic.
Thin wrappers closed such gaps in the adapted public skills (\autoref{sec:eval:compile}), so skill authors and tool designers can build enforceability in with helpers that return receipts with a digest, approvals through native questions, and a completion acknowledgment.

\noindent\textbf{Why no model at runtime.}
A monitor that asks a model about each request would read the same untrusted task content as the agent.
Injected text could then sway it~\cite{DBLP:conf/naacl/ZhanFPK25,shi2024optimization,ruan2026heimdallrcharacterizingdetectingllminduced}, no guarantee like \autoref{thm:safety} would hold, and every request would cost a model call.

\noindent\textbf{Scope and validity.}
\tool assumes a trusted \texttt{SourceSkill} (\textbf{M5}), and screening malicious skills~\cite{liu_malicious_2026,schmotz_skill-inject_2026} or MCP servers~\cite{zhao2025mcpserversattacktaxonomy} is orthogonal.
The guarantees concern boundary events relative to $C$.
They carry over to the skill under two per-skill assumptions.
Each helper checks the identity of the object it acts on (\textbf{M3}), and an audit confirms that $C$ matches the prose and that no helper hides a protected step (\textbf{M5}).
The audit is needed because a contract can pass acceptance and still be stricter than its source, as with the pinned plan ID and the repeated key check of speech (\autoref{sec:eval:compile}).
\ac{skillcfg} assumes serialized single-skill sessions (\textbf{M4}), its guards read only the latest outcome and counters, and its value relations are limited to equality, disequality, and single question templates.
Concurrent dispatch, multi-skill composition, and richer predicates remain future work.

\section{Related Work}
\label{sec:related-work}

\noindent\textbf{CFI and system-call monitoring.}
\tool keeps the policy of \ac{cfi}~\cite{abadi_control-flow_2009} and adopts the mechanism of system-call monitors, which check a process against a model of its calls at the kernel boundary~\cite{wagner2000intrusion,sekar2000fast} and face mimicry by call sequences that the model accepts~\cite{wagner2002mimicry}.
Unlike models extracted from code, \tool's contract is synthesized from prose and verified before use; moreover, each transition depends on correlated call results because the proposer itself is untrusted.

\noindent\textbf{Enforceability and workflow monitoring.}
Execution-monitor theory characterizes which policies a monitor can enforce~\cite{schneider2000enforceable}, including the roles of merely observable actions~\cite{basin2013enforceable} and of action results~\cite{ligatti2010theory}.
\autoref{prop:obs-limit} applies this reasoning to a declared boundary and asks which obligations stay enforceable when the agent can move a step out of view.
Reference monitors for workflow systems and execution-history access control likewise start from explicit task definitions~\cite{crampton2005reference,fong2004access}, while \tool derives its contract from a trusted skill written in natural language.

\noindent\textbf{Securing LLM agents.}
Most agent defenses~\cite{kim2026sok} protect the user's task from prompt injection.
They isolate untrusted components~\cite{wu2025isolategpt}, fix the plan before untrusted data is read~\cite{debenedetti2026camel,DBLP:conf/ndss/LiMRRON26}, learn behavioral bounds from benign runs~\cite{liu_traceaegis_2025,sequeira_agent-sentry_2026}, or constrain multi-agent invocations with CFI-style graphs~\cite{jha2026breaking}.
Privilege mechanisms bound which operations a tool call~\cite{shi_progent_2025} or a skill~\cite{wu2026skillscope,pan2026skillguard} may use, which alone does not exclude deviations \ding{172}--\ding{175} of \autoref{sec:threat-security}.

\begin{table}[t]
  \centering
  \caption{Closest enforcement systems. Evidence advances the monitor's state. Pre-disp.\ marks blocking before dispatch. Unchecked is how a requirement the monitor cannot check is treated (assumed observable, or rejected at compile time if mandatory). Agent marks whether the native agent keeps control. --- means not addressed.}
  \label{tab:related}
  \scriptsize
  \setlength{\tabcolsep}{2.4pt}
  \begin{tabularx}{\linewidth}{@{}l>{\raggedright\arraybackslash}Xllll@{}}
    \toprule
    \textbf{System} & \textbf{Policy source} & \textbf{Evidence} & \textbf{Pre-disp.} & \textbf{Unchecked} & \textbf{Agent} \\
    \midrule
    Progent~\cite{shi_progent_2025} & LLM, updated in run & --- & \ding{51} & --- & \ding{51} \\
    AgentSpec~\cite{wang2026agentspec} & user or LLM rules & trajectory & \ding{51} & --- & \ding{51} \\
    FORGE~\cite{palumbo_policy_2026} & fixed Datalog & event graph & \ding{51} & assumed & \ding{51} \\
    VIGIL~\cite{li2026vigil} & LLM drafts per run & trace facts & partly & reported & \ding{51} \\
    SpecHarness~\cite{li2026holds} & compiled per task & verifiers & partly & advisory & \ding{51} \\
    SkillSentry~\cite{lu2026skillsentry} & LLM, self-evolving & proposals & \ding{51} & admitted & \ding{51} \\
    SIGIL~\cite{dantanarayana2026sigil} & compiled harness & harness & n/a & --- & \ding{55} \\
    \tool & skill, compiled once & results & \ding{51} & rejected & \ding{51} \\
    \bottomrule
  \end{tabularx}
\end{table}

\noindent\textbf{Enforcing specifications and skills.}
Closest to \tool are systems that enforce a given specification or skill procedure (\autoref{tab:related}).
Among policy monitors, AgentSpec checks user-written rules before each action~\cite{wang2026agentspec}, FORGE evaluates Datalog policies over an event graph~\cite{palumbo_policy_2026}, VIGIL checks temporal and value-flow policies that an \ac{llm} drafts for each run~\cite{li2026vigil}, and SpecHarness tracks obligations whose state only qualified evidence can establish~\cite{li2026holds}.
SpecHarness audits which action surfaces are closed, and its evidence authority and closure audit resemble our two closure principles, but it leaves the remaining requirements advisory or defers them to post-execution validation.
SIGIL lowers a skill into a harness that takes execution away from the agent~\cite{dantanarayana2026sigil}, and SkillSentry checks the agent's actions through hooks but credits progress from its proposals~\cite{lu2026skillsentry}.
None of these systems characterizes which obligations remain enforceable when the agent itself decides whether a step runs in its reasoning, a mediated call, or generated code.
\tool starts from this question, and its contract is compiled once from a trusted skill, verified for deterministic admission, rendered together with the agent's instructions, and checked before dispatch.

\section{Conclusion and Outlook}
\label{sec:conclusion}

When an agent skill moves a procedure into prompt text, its control skeleton survives, but only obligations whose effects and evidence cross the tool-call boundary can be enforced without trusting the model.
\tool confines \ac{llm} interpretation to compile time and admits a tool request only on trusted evidence. In our evaluation, it blocked violations that an allowlist admits while preserving legitimate workflows.
As computation moves into models, the reference-monitor principles of earlier systems still apply once the boundary and its evidence are explicit.
Looking ahead, extending enforcement to $R_3$ requires verifying that a generated script matches its task and confining its execution with \ac{cfi}.
Steps now in $R_1$ become enforceable only when tools and hosts expose trusted evidence at the boundary.

\bibliographystyle{IEEEtran}
\bibliography{references}

\appendices

\section{Notation}
\label{sec:appendix-notation}

\autoref{tab:notation-full} lists the notation of the paper.

\begin{table*}[t]
  \centering
  \caption{Notation summary.}
  \label{tab:notation-full}
  \footnotesize
  \setlength{\tabcolsep}{3pt}
  \renewcommand{\arraystretch}{1.05}
  \begin{minipage}[t]{0.495\textwidth}
  \vspace{0pt}
  \begin{tabularx}{\linewidth}{@{}l >{\raggedright\arraybackslash}X@{}}
    \toprule
    \textbf{Symbol} & \textbf{Meaning} \\
    \midrule
    \multicolumn{2}{@{}l}{\emph{Execution model (\autoref{sec:execution-model})}} \\
    $\tau$; $\pi_B(\tau) = \tau_B$ & Full trace; its projection onto boundary $B$ \\
    $\mathsf{req}\langle i, \iota, a\rangle$ & Request for interface $i$ with call ID $\iota$ and arguments $a$ \\
    $\mathsf{res}\langle \iota, v, F\rangle$ & Result of call $\iota$: outcome $v$, captures $F$ \\
    $R_1$, $R_2$, $R_{3\mathrm{a}}$, $R_{3\mathrm{b}}$ & Internal reasoning; mediated calls; trusted helper; model-written script \\
    $\varphi$, $\varphi^+$ & Safety property; variant requiring a pass result at $B$ \\
    $D_\varphi(\tau, r)$ & Required decision on request $r$ (must-allow / must-refuse / free) \\
    $M$ & $B$-monitor (a function of $\pi_B(\tau)$ and $r$) \\
    \midrule
    \multicolumn{2}{@{}l}{\emph{Catalog and contract (\autoref{sec:design:contract})}} \\
    $\mathit{Cat} = (I, \mathcal{R}, \mathcal{X})$ & Interfaces, value relations, capture registers \\
    $L(i)$; $\mathit{dec}_i$; $O_i$ & Request language; result decoder; outcomes (e.g., $\{\mathsf{pass}, \mathsf{fail}\}$) \\
    $C$ & SkillCFG $(Q, q_0, \mathit{Occ}, E, \mathit{Ctr}, \mathit{Exit})$ \\
    $o = (i_o, \mathit{Rel}_o, \mathit{cap}_o)$ & Occurrence of $i_o$ (e.g., $\mathtt{o\_edit}$, $\mathtt{o\_repair}$ of $\mathtt{i\_edit}$) \\
    $\mathit{Occ}(i)$ & Occurrences of interface $i$ \\
    $\mathsf{req}(o)$, $\mathsf{res}(o)$ & Request and result nodes of $o$ \\
    $Q_{\mathrm{idle}}$ & Idle locations: $q_0$ and all $\mathsf{res}(o)$ \\
    $q \xrightarrow{g/u} \mathsf{req}(o)$ & Edge with guard $g(\mathit{fact}, \mathit{ctr})$ and counter update $u$ \\
    $\mathit{Exit}$; $E_{\mathit{Exit}}$ & $\{\mathsf{success}, \mathsf{failure}, \mathsf{abort}\}$; exit edges \\
    $\kappa = (q, \mathit{fact}, \mathit{ctr}, X)$ & Configuration: location, last outcome, counters, registers \\
    $\mathsf{req}(o, a)$, $\mathsf{res}(o, v, F)$ & Contract steps labeled by occurrence \\
    $\operatorname{Paths}(C)$; $\operatorname{Run}_C$ & Permitted step sequences; configuration reached \\
    \textbf{C}/\textbf{S}/\textbf{R}/\textbf{F} & Integrity, interface, relation, flow checks \\
    \bottomrule
  \end{tabularx}
  \end{minipage}\hfill
  \begin{minipage}[t]{0.495\textwidth}
  \vspace{0pt}
  \begin{tabularx}{\linewidth}{@{}l >{\raggedright\arraybackslash}X@{}}
    \toprule
    \textbf{Symbol} & \textbf{Meaning} \\
    \midrule
    \multicolumn{2}{@{}l}{\emph{Compiler (\autoref{sec:appro-compiler}, \autoref{alg:compiler})}} \\
    $S$, $H$; $\mathit{Obl}$; $K$ & Source skill, host capabilities; obligations; SkillSpec \\
    $\sigma^\# = (q, \mathit{fact}, \mathit{ctr}, \Gamma)$ & Abstract configuration \\
    $\Gamma = (D, \sim, \mathit{Neq})$ & Bound registers, equalities over $D$ and literals $\mathit{Lit}$, disequalities \\
    $\gamma(\sigma^\#)$ & Concrete configurations consistent with $\sigma^\#$ \\
    $W$, $\mathit{Seen}$, $B_{\max}$ & Worklist, visited states, state budget \\
    $E^{\mathit{en}}_{\sigma^\#}(o)$; $\mathit{En}$ & Enabled edges into $\mathsf{req}(o)$; occurrences with such an edge \\
    $\operatorname{refs}(\mathit{Rel}_o)$ & Registers referenced by $\mathit{Rel}_o$ \\
    $\mathit{Cand}$ & Enabled occurrences with satisfiable relations \\
    $\Gamma \models \exists a.\,\mathit{Rel}_o$ & Every valuation consistent with $\Gamma$ admits arguments satisfying $\mathit{Rel}_o$ \\
    $\mathsf{Terminal}(\sigma^\#)$ & $\sigma^\#$ enables an exit; not explored further \\
    $\mathsf{Sat}$ & Satisfiability check for $=$/$\neq$ constraints \\
    $\operatorname{Sig}(o)$ & Handler signature (decoder, captures, update, successors) \\
    \texttt{P1}--\texttt{P7} & Acceptance properties (\autoref{tab:acceptance-properties}) \\
    \midrule
    \multicolumn{2}{@{}l}{\emph{Runtime (\autoref{sec:appro-runtime}, \autoref{alg:runtime})}} \\
    $\sigma$ & $\kappa$ extended with $\mathit{pending}$ and $\mathit{term}$ \\
    $\mathit{pending}$; $\mathit{term}$ & In-flight call lock; entered exit or $\bot$ \\
    $\mathit{Plans}$, $p^*$ & Candidate plans; selected plan \\
    $p = \langle o, u, \mathit{inv}, \mathit{reqCap} \rangle$ & Plan: occurrence, counter update, invalidated registers, request captures \\
    $X \oplus Y$ & Override of valuation $X$ by $Y$ \\
    $\tau_B^{\mathrm{adm}}$ & Admitted workflow requests and accepted results \\
    $\varphi_C$ & Property induced by contract $C$ (\autoref{prop:transparency}) \\
    $\delta$ & Refusal payload $\langle \mathit{code}, \mathit{layer}, \mathit{missing}, \mathit{next}\rangle$ \\
    \textbf{M1}--\textbf{M5} & System and trust assumptions (\autoref{sec:threat-model}) \\
    \bottomrule
  \end{tabularx}
  \end{minipage}
\end{table*}

\section{Compiler and Enforcer Details}
\label{sec:appendix-algorithms}

\autoref{alg:compiler} gives the compilation procedure of \autoref{sec:appro-compiler}.
Below we define its subroutines and those of \autoref{alg:runtime}.

\setcounter{algocf}{0}%
\begin{algorithm}[!b]
\footnotesize
\caption{Enforceability-Aware Propose-and-Check Compilation}
\label{alg:compiler}
\KwInput{Source skill $S$, host capabilities $H$}
\KwOutput{Verified $\mathit{SkillBundle}$ or $\mathsf{Unsupported}(\mathit{diag})$}
$\mathit{Cat} \leftarrow \mathsf{LLM\_Catalog}(S, H)$;\ \textbf{if} $\neg \mathsf{VerifyInterfaces}(\mathit{Cat})$ \textbf{then} \Return{$\mathsf{Unsupported}(\texttt{P1}, \texttt{P2})$}\;
$\mathit{Obl} \leftarrow \mathsf{LLM\_Obligations}(S, \mathit{Cat})$;\ \textbf{if} $\exists\, \mathsf{unsupported} \in \mathit{Obl}$ \textbf{then} \Return{$\mathsf{Unsupported}(\text{Prop.~\ref{prop:obs-limit}})$}\;
$K \leftarrow \mathsf{LLM\_Synthesize}(S, \mathit{Cat}, \mathit{Obl})$;\ \textbf{if} $\neg \mathsf{Agree}(K, \mathit{Obl})$ \textbf{then} \Return{$\mathsf{Unsupported}(\mathit{diag})$}\;
$C \leftarrow \mathsf{Lower}(K, \mathit{Cat})$;\ $W, \mathit{Seen} \leftarrow \{\sigma^\#_0\}$ \tcp*{$\sigma^\# = (q, \mathit{fact}, \mathit{ctr}, \Gamma)$}
\While{$W \neq \emptyset$}{
    $\sigma^\# = (q, \mathit{fact}, \mathit{ctr}, \Gamma) \leftarrow \mathsf{dequeue}(W)$\;
    $\mathit{En} \leftarrow \{o \in \mathit{Occ} \mid E^{\mathit{en}}_{\sigma^\#}(o) \neq \emptyset\}$;\ $\mathit{Cand} \leftarrow \{o \in \mathit{En} \mid \mathsf{Sat}(\Gamma \land \mathit{Rel}_o)\}$\;
    \Require{$\forall o \in \mathit{En} : \operatorname{refs}(\mathit{Rel}_o) \subseteq D \land \lvert E^{\mathit{en}}_{\sigma^\#}(o) \rvert = 1$;\ $\exists o \in \mathit{Cand} : \Gamma \models \exists a.\,\mathit{Rel}_o$;\ $\forall o, o' \in \mathit{Cand} : i_o = i_{o'} \land \mathsf{Sat}(\Gamma \land \mathit{Rel}_o \land \mathit{Rel}_{o'}) \Rightarrow \operatorname{Sig}(o) = \operatorname{Sig}(o')$} \tcp*{\texttt{P3}\textrm{--}\texttt{P5}}
    \For{$o \in \mathit{Cand}$, $v \in O_{i_o}$}{
        $\sigma^{\#\prime} \leftarrow \mathsf{AbstractPost}(\sigma^\#, o, v)$;\ \Require{$\mathsf{ValidContinuation}(\sigma^{\#\prime})$} \tcp*{\texttt{P4, P7}}
        \textbf{if} $\sigma^{\#\prime} \notin \mathit{Seen}$ \textbf{then} $\mathit{Seen} \leftarrow \mathit{Seen} \cup \{\sigma^{\#\prime}\}$;\ \textbf{if} $\neg\mathsf{Terminal}(\sigma^{\#\prime})$ \textbf{then} $\mathsf{enqueue}(W, \sigma^{\#\prime})$\;
    }
    \textbf{if} $\lvert \mathit{Seen} \rvert > B_{\max}$ \textbf{then} \Return{$\mathsf{Unsupported}(\mathit{diag})$}\;
}
\Require{$\mathsf{AllVisited}(\mathit{Occ}, \mathit{Seen})$} \tcp*{\texttt{P6}}
\Return{$\mathsf{Package}(\mathsf{RenderAgentSkill}(K, \mathit{Cat}), C, \mathit{Cat}, S)$}\;
\end{algorithm}

\noindent\textbf{Proposal stages.}
The catalog stage returns $\mathit{Cat}$ (\autoref{def:catalog}).
The obligation stage returns a plan $\mathit{Obl}$ of cited source obligations, each classified as \texttt{enforce}, \texttt{guidance}, or \texttt{unsupported}, and assigns each enforced obligation workflow occurrences of catalog interfaces that cover every outcome.
The synthesis stage returns $K$, which declares these occurrences with their base interfaces, cited obligations, value relations, and capture plans, together with counters, guarded transitions, exits, and the text of the guidance obligations, which $\mathsf{RenderAgentSkill}$ copies into \texttt{AgentSkill}.
Each stage is one model call without retry, and a failed check ends compilation.

\noindent\textbf{Checks and lowering.}
$\mathsf{VerifyInterfaces}$ checks \texttt{P1} and \texttt{P2}.
$\mathsf{Agree}(K, \mathit{Obl})$ holds iff $K$ declares exactly the occurrences of $\mathit{Obl}$, none omitted, duplicated, or added, each with the base interface and source obligation that $\mathit{Obl}$ fixes.
$\mathsf{Lower}$ maps each occurrence to its nodes $\mathsf{req}(o)$ and $\mathsf{res}(o)$ joined by the correlation edge, each transition of $K$ to a request edge with its guard and counter update, and each exit of $K$ to an exit edge, preserving occurrence and obligation identities.

\noindent\textbf{Exploration.}
$E^{\mathit{en}}_{\sigma^\#}(o)$ is the set of enabled edges into $\mathsf{req}(o)$, $D$ the registers bound in $\Gamma$, and $\operatorname{refs}(\mathit{Rel}_o)$ those referenced by $\mathit{Rel}_o$.

$\mathsf{AbstractPost}(\sigma^\#, o, v)$ applies the request and result steps of $o$ to $\sigma^\#$. It removes from $\Gamma$ the registers that admitting $o$ clears, namely those that $\mathit{cap}_o$ invalidates and those that a capture of $o$ may write on the request or on any result, and keeps the constraints among the remaining registers and literals. It then adds the request captures with the constraints that $\mathit{Rel}_o$ imposes on them before the removal, adds a fresh unconstrained class for each result capture that $\mathit{cap}_o(v)$ publishes, and sets $(q, \mathit{fact}, \mathit{ctr})$ to $(\mathsf{res}(o), v, u(\mathit{ctr}))$ for the update $u$ of the unique enabled edge into $\mathsf{req}(o)$.

$\mathsf{ValidContinuation}(\sigma^{\#\prime})$ holds iff $\sigma^{\#\prime}$ enables at most one exit edge and, if it enables one, no request edge (\texttt{P7}). A successor that enables no exit is non-terminal and must meet the progress condition of Line~8 of \autoref{alg:compiler} when dequeued (\texttt{P4}).
A successor is $\mathsf{Terminal}$ iff it enables an exit edge. Such a successor is recorded for coverage but not explored because the runtime refuses every request once $\mathit{term} \neq \bot$.

\noindent\textbf{Coverage and packaging.}
$\mathsf{AllVisited}(\mathit{Occ}, \mathit{Seen})$ holds iff, for every $o \in \mathit{Occ}$ and $v \in O_{i_o}$, some state in $\mathit{Seen}$ has location $\mathsf{res}(o)$ and $\mathit{fact} = v$ (\texttt{P6}).
Coverage is over the explored abstract states. When several occurrences are always enabled together with equal signatures, the runtime credits only the one with the least identifier, which changes no decision (Lemma~4, \autoref{sec:appendix-sketches}).
The state budget $B_{\max}$ is 4096.
$\mathsf{Package}$ assembles \texttt{AgentSkill}, $\mathit{Cat}$, $C$, and the package members of $S$ into the \texttt{SkillBundle} with a manifest of their digests (\textbf{M5}).

\noindent\textbf{Why not track candidate states.}
An operation invoked from several points with different continuations or permissions does not identify the control state~\cite{sekar2000fast}.
For register contracts, a monitor that merges candidate states also merges their permissions, and their valuations conflict when one candidate invalidates a register that another keeps.
When two prerequisite steps share an interface, one invocation could also count as both steps.
\texttt{P3} (\autoref{tab:acceptance-properties}) rules out these cases at compile time.

\noindent\textbf{Enforcer subroutines.}
$\mathsf{ResolveBaseInterface}(\mathit{tool}, a, I)$ parses $r$ under the closed argument schemas of $I$ (\textbf{M3}) and returns the unique $i$ with $a \in L(i)$ (\texttt{P1}), or $\bot$ if there is none.
$\mathsf{AuxiliaryOrRefuse}(r)$ admits $r$ as an auxiliary operation if a deployment-fixed adapter matches it and otherwise refuses it as an unexpected action.
The adapters are disjoint from $I$ and cover bounded reads of inventoried package files and of scoped task files, and bounded directory listings.
Since an auxiliary call supplies no capture, a read whose result a relation uses must be a workflow interface.
$\mathsf{BuildPlan}(o, a, \sigma)$ returns the plan $p = \langle o, u, \mathit{inv}, \mathit{reqCap} \rangle$, where $u$ is the counter update of the unique enabled edge into $\mathsf{req}(o)$, $\mathit{inv}$ the registers that $\mathit{cap}_o$ invalidates or that a result capture of $o$ may write, and $\mathit{reqCap}$ the request captures of $\mathit{cap}_o$ evaluated on $a$.
$\mathsf{SelectCanonical}$ returns the plan with the least occurrence identifier.
$\mathsf{ReleaseOnce}(a)$ hands $a$ to the host for exactly one dispatch under call ID $\iota$ (\textbf{M1}).
$\mathsf{Filter}(F, o, v)$ restricts $F$ to the registers that $\mathit{cap}_o(v)$ publishes, all of which a decoded receipt carries, since a receipt that lacks one is incomplete (\textbf{M3}).
$\mathsf{Halt}$ stops workflow progression and keeps $\mathit{pending}$ set, so Line~2 of \autoref{alg:runtime} refuses every later request, also after a failed auxiliary read.
$\mathsf{Refuse}(\mathit{code})$ dispatches nothing and returns the payload $\delta$ (\autoref{sec:design:guidance}) with that code. The code $\mathsf{Integrity\_C} \mid \mathsf{Flow\_F}$ denotes the first code if $\mathit{pending} \neq \bot$ and the second if $\mathit{term} \neq \bot$, and $\mathsf{InvalidTransition\_F\_R}$ reports layer \textbf{F} if no occurrence in $\mathit{Occ}(i)$ has an enabled entering edge and \textbf{R} otherwise.

\noindent\textbf{Halting on auxiliary reads.}
$\mathsf{Halt}$ keeps $\mathit{pending}$ set also when the failed result belongs to an auxiliary read.
A read outside the adapters' scope or of a missing or oversized file is refused before dispatch, so an auxiliary read halts only when its result contradicts the admitted file.

\section{Compilation Example}
\label{sec:appendix-example}

\autoref{tab:compile-trace} traces three requirements of the document-editing skill through one compilation (\autoref{sec:eval:setup}).
Identifiers are those of the compiled bundle, which refines \autoref{fig:contract-automaton}: $\mathtt{validatedHash}$ plays the role of $\mathtt{draft\_sha}$, and each repair cycle re-renders and re-validates the working copy.

\begin{table*}[t]
  \centering
  \caption{Three source requirements of the document-editing skill through the compiled artifacts; $\to$ is a transition of $K$.}
  \label{tab:compile-trace}
  \footnotesize
  \setlength{\tabcolsep}{3pt}
  \renewcommand{\arraystretch}{1.1}
  \begin{tabularx}{\textwidth}{@{}>{\raggedright\arraybackslash}p{0.2\textwidth} >{\raggedright\arraybackslash}p{0.13\textwidth} >{\raggedright\arraybackslash}X >{\raggedright\arraybackslash}p{0.25\textwidth}@{}}
    \toprule
    \textbf{Source (\texttt{SKILL.md})} & \textbf{$\mathit{Obl}$} & \textbf{$K$, $C$, and $\mathit{Cat}$} & \textbf{\texttt{AgentSkill}} \\
    \midrule
    ``At most three repair cycles after the initial validation are permitted.''
    & \texttt{enforce}
    & Counter $\mathit{repairs} \le 3$; \texttt{validate-repair} $\to$ \texttt{repair-abstract} with guard $\mathsf{fail} \land \mathit{repairs} < 3$ and update $\mathit{repairs} \mathrel{+}= 1$; \texttt{validate-repair} $\to$ \texttt{acknowledge-failure} with guard $\mathsf{fail} \land \mathit{repairs} = 3$.
    & \texttt{If the observed outcome is "fail" and "repairs" is below 3, go to step 6 (repairs += 1 when that next call is admitted).} \\
    \midrule
    ``permission to overwrite the source authorizes only the \texttt{replace} command.''
    & \texttt{enforce}
    & \texttt{ask-delivery} $\to$ \texttt{deliver-replacement} with guard $\mathsf{replace}$; relation $a[\texttt{--candidate-sha256}] = X[\mathtt{validatedHash}]$, where $\mathtt{validatedHash}$ is captured only from a $\mathsf{pass}$ result of validation.
    & \texttt{Require actual argument '--candidate-sha256' to equal the latest available observed value <reuse:validatedHash>.} \\
    \midrule
    ``describe the change in the median''
    & \texttt{guidance}
    & No occurrence, guard, or relation.
    & Rendered as instruction text; not enforced. \\
    \bottomrule
  \end{tabularx}
\end{table*}

\section{Formal Proofs}
\label{sec:appendix-proofs}

This appendix proves \autoref{prop:obs-limit}, \autoref{thm:safety}, \autoref{thm:runtime-determinism}, and \autoref{prop:transparency}.

\subsection{Proof of \autoref{prop:obs-limit}}
\label{sec:appendix-prop}

\begin{IEEEproof}
\emph{Only if:} Suppose such $\tau_1, \tau_2$, and $r$ exist. A $B$-monitor $M$ receives the same input $(\pi_B(\tau_1), r) = (\pi_B(\tau_2), r)$ on both prefixes and must return the same decision; $\mathsf{allow}$ violates soundness on $\tau_1$, and $\mathsf{refuse}$ violates transparency on $\tau_2$.
\emph{If:} Otherwise, for every projection $h$ and request $r$, the prefixes with $\pi_B(\tau) = h$ never demand both $\mathsf{must\text{-}refuse}$ and $\mathsf{must\text{-}allow}$. The monitor that returns $\mathsf{refuse}$ exactly when some such prefix demands $\mathsf{must\text{-}refuse}$ is therefore sound and transparent.
\end{IEEEproof}

\subsection{Contract Step Semantics}
\label{sec:appendix-semantics}

Let $\kappa = (q, \mathit{fact}, \mathit{ctr}, X)$ be an idle configuration of $C$.
\begin{itemize}[leftmargin=*]
    \item \emph{Request step $\mathsf{req}(o, a)$:} enabled at $\kappa$ when exactly one edge $q \xrightarrow{g / u} \mathsf{req}(o)$ satisfies $g(\mathit{fact}, \mathit{ctr}) = \mathsf{true}$, $a \in L(i_o)$, all captures referenced by $\mathit{Rel}_o$ are in $\operatorname{dom}(X)$, and $\mathit{Rel}_o(a, X)$ holds. 
    Firing it sets $\mathit{ctr} \leftarrow u(\mathit{ctr})$ and $\mathit{fact} \leftarrow \bot$, removes from $X$ every register that $\mathit{cap}_o$ invalidates or that a capture of $o$ may write (on the request or any result), records the request captures from $a$, and moves control to $\mathsf{req}(o)$.
    \item \emph{Result step $\mathsf{res}(o, v, F)$:} enabled at $\mathsf{req}(o)$ when the correlated return decodes to $v \in O_{i_o}$ with extracted fields $F$. Firing it sets $\mathit{fact} \leftarrow v$, publishes into $X$ the captures in $F$ designated by $\mathit{cap}_o(v)$, moves control to $\mathsf{res}(o)$, and enters exit $t \in \mathit{Exit}$ if exactly one exit edge $\mathsf{res}(o) \xrightarrow{g} t$ is enabled; \texttt{P7} further guarantees that at most one is enabled and that it excludes every request edge.
    A configuration that has entered an exit is \emph{terminal} and enables no step.
\end{itemize}

\subsection{Proof of \autoref{thm:safety}}

\begin{IEEEproof}
For a runtime state $\sigma = (q, \mathit{pending}, \mathit{fact}, \mathit{ctr}, X, \mathit{term})$, let $\kappa(\sigma) = (q, \mathit{fact}, \mathit{ctr}, X)$ denote its contract configuration (\autoref{def:contract}).
The slot $\mathit{pending}$ holds either a workflow entry $\langle \iota, p^*, a, \mathit{dec}_i \rangle$ or an auxiliary entry $\langle \iota, \mathsf{aux}, a, \mathit{dec}_{\mathsf{aux}} \rangle$ (\autoref{sec:appro-runtime}).
Line numbers in this proof refer to \autoref{alg:runtime}.
We proceed by induction over the atomic monitor transactions and maintain three invariants at every prefix:
\begin{enumerate}[label=(\roman*), leftmargin=*]
    \item \emph{Trace inclusion:} $\tau_B^{\mathrm{adm}} \in \operatorname{Paths}(C)$ and $\kappa(\sigma) = \operatorname{Run}_C(\tau_B^{\mathrm{adm}})$.
    \item \emph{Lock and closure:} $\mathit{pending}$ holds a workflow entry iff $\tau_B^{\mathrm{adm}}$ ends with an unmatched request step $\mathsf{req}(o, a)$, in which case $q = \mathsf{req}(o)$ and $p^*.o = o$; otherwise $q$ is an idle location ($q_0$ or some $\mathsf{res}(o)$). Moreover, $\mathit{term} = t \neq \bot$ iff $\tau_B^{\mathrm{adm}}$ ends at $\mathsf{res}(o)$ with a unique enabled exit edge $\mathsf{res}(o) \xrightarrow{g} t$.
    \item \emph{Provenance:} every binding in $X$ was recorded from the arguments of an admitted request ($p^*.\mathit{reqCap}$) or from a successfully decoded workflow result ($\mathsf{Filter}(F, p^*.o, v)$) in this session, and no later admitted occurrence has invalidated or overwritten it.
\end{enumerate}

\emph{Base case.}
On verified activation (\textbf{M5}), $\tau_B^{\mathrm{adm}} = \varepsilon$, $\kappa(\sigma_0) = \kappa_0 = \operatorname{Run}_C(\varepsilon)$, $\mathit{pending} = \bot$, and $\mathit{term} = \bot$, so (i)--(iii) hold.
Steps in $R_1$ emit no boundary events and do not change $\sigma$.

\emph{Inductive step.}
By \textbf{M1}, every native request reaches $\mathsf{OnRequest}$ before execution; by \textbf{M4}, $\sigma$ changes only through the following transactions.
\begin{itemize}[leftmargin=*]
    \item \emph{Case 1: refusal.}
    Every refusal (Lines~2, 3, 5, and~6) is decided before the commit at Line~8 and the release at Line~9.
    No handler in $R_2/R_3$ runs, $\tau_B^{\mathrm{adm}}$ is unchanged, and $\sigma' = \sigma$. This proves Part~(a) and preserves (i)--(iii).

    \item \emph{Case 2: workflow admission of $p^*$ with $o = p^*.o$.}
    Line~2 gives $\mathit{pending} = \bot$ and $\mathit{term} = \bot$, so by (ii) $q$ is a non-terminal idle location. Line~3 returns the interface $i$ with $a \in L(i)$, unique by \texttt{P1}; since $o \in \mathit{Occ}(i)$, $a \in L(i_o)$. Line~4 checks that exactly one edge $q \xrightarrow{g/u} \mathsf{req}(o)$ satisfies $g(\mathit{fact}, \mathit{ctr})$ and that $\mathit{Rel}_o(a, X)$ holds; a relation that references an unbound register evaluates to false, so all referenced registers are in $\operatorname{dom}(X)$.
    By (iii), these values were recorded from earlier admitted request arguments or verified result captures that remain valid; model claims in $R_1$ and untrusted content in $R_2/R_3$ can influence them only through admitted arguments, which satisfied the relations of their own occurrence, or through receipts of trusted helpers (\textbf{M3}). This proves Part~(c).
    Hence $\mathsf{req}(o, a)$ is an enabled request step of $C$ at $\kappa(\sigma)$.
    Line~8 applies exactly the request-step update of \autoref{sec:appendix-semantics} and pins a workflow entry in $\mathit{pending}$, and Line~9 dispatches $a$ once (\textbf{M1}).
    Thus $\tau_B^{\mathrm{adm}} \cdot \mathsf{req}(o, a) \in \operatorname{Paths}(C)$, and (i)--(iii) hold.

    \item \emph{Case 3: auxiliary admission.}
    Line~2 gives $\mathit{pending} = \bot$ and $\mathit{term} = \bot$, and Line~3 gives $i = \bot$, so the request matches no workflow interface.
    If a deployment-fixed auxiliary adapter matches it, the enforcer pins an auxiliary entry in $\mathit{pending}$ and dispatches it once; otherwise the request is refused as in Case~1.
    $\kappa(\sigma)$ and $\tau_B^{\mathrm{adm}}$ do not change, and by \textbf{M3} the call does not modify protected targets, so (i)--(iii) hold.

    \item \emph{Case 4: result of a workflow call.}
    By Line~11 and \textbf{M2}, the result is accepted only if its call ID and dispatched arguments match the pinned workflow entry, and the enforcer halts otherwise.
    By (ii), $\tau_B^{\mathrm{adm}}$ ends with $\mathsf{req}(o, a)$.
    If $\mathit{dec}_i$ fails (Line~12), the enforcer halts without clearing $\mathit{pending}$, so Line~2 refuses every later request and $\tau_B^{\mathrm{adm}}$ is unchanged.
    Otherwise, by \texttt{P2} and \textbf{M3}, $\mathit{dec}_i$ yields a unique $(v, F)$ with $v \in O_{i_o}$.
    The edge $\mathsf{req}(o) \to \mathsf{res}(o)$ is unconditional, so $\mathsf{res}(o, v, F)$ is an enabled result step. Line~13 applies its update and clears $\mathit{pending}$, and Line~14 sets $\mathit{term}$ exactly when a unique exit edge is enabled.
    Thus $\tau_B^{\mathrm{adm}} \cdot \mathsf{res}(o, v, F) \in \operatorname{Paths}(C)$, and (i)--(iii) hold.

    \item \emph{Case 5: result of an auxiliary call.}
    The result must match the pinned auxiliary entry and decode under $\mathit{dec}_{\mathsf{aux}}$; otherwise the enforcer halts.
    On success, only $\mathit{pending}$ is cleared (Line~12); $\kappa(\sigma)$ and $\tau_B^{\mathrm{adm}}$ do not change, and the result supplies no outcome, capture, or approval.
\end{itemize}
Each call identifier $\iota$ is unique (\textbf{M2}) and is admitted at most once, and its single result clears the slot that pinned it.
Hence each native invocation contributes at most one request and one result to $\tau_B^{\mathrm{adm}}$. Invariant~(i) at every prefix proves Part~(b), and invariant~(ii) with Line~2 proves Part~(d).
\end{IEEEproof}

\subsection{Proof of \autoref{thm:runtime-determinism}}
\label{sec:appendix-sketches}

\noindent\textbf{Abstract domain.}
\autoref{alg:compiler} explores abstract configurations $\sigma^\# = (q, \mathit{fact}, \mathit{ctr}, \Gamma)$ with $\Gamma = (D, \sim, \mathit{Neq})$, where $D \subseteq \mathcal{X}$ is the set of bound registers, $\sim$ is an equivalence relation over $D \cup \mathit{Lit}$ for the finite set $\mathit{Lit}$ of contract literals, and $\mathit{Neq}$ is a symmetric, irreflexive relation over $\sim$-classes.
Unqualified line numbers in this subsection refer to \autoref{alg:compiler}.
An idle configuration $\kappa = (q', \mathit{fact}', \mathit{ctr}', X)$ is in $\gamma(\sigma^\#)$ iff $(q', \mathit{fact}', \mathit{ctr}') = (q, \mathit{fact}, \mathit{ctr})$, $\operatorname{dom}(X) = D$, and $X$ respects $\sim$ and $\mathit{Neq}$.
Respecting is one-directional: $u \sim v$ implies $X(u) = X(v)$, and classes that $\mathit{Neq}$ separates hold different values, while unrelated classes may hold equal values. A fresh class for a result capture therefore covers every value the result can carry, including one that equals an existing register.

\noindent\textbf{Lemma 1 (Termination).}
$Q$, $\bigcup_i O_i$, and $D \cup \mathit{Lit}$ are finite, so for bounded counters there are finitely many canonical $\sigma^\#$.
Otherwise, the budget check at Line~12 of \autoref{alg:compiler} terminates exploration with $\mathsf{Unsupported}$.

\noindent\textbf{Lemma 2 (Exactness of $\mathsf{Sat}$).}
Let $\phi$ be $\Gamma \land \mathit{Rel}_o$ or $\Gamma \land \mathit{Rel}_o \land \mathit{Rel}_{o'}$.
$\mathsf{Sat}(\phi)$ merges $\sim$-classes under the equalities of $\phi$, adds the disequalities of $\phi$ to $\mathit{Neq}$, and returns false iff some class becomes unequal to itself or two distinct literals fall into the same class.
\emph{Soundness:} if some valuation $X$ of a configuration in $\gamma(\sigma^\#)$ and some arguments $a$ satisfy $\phi$, no such conflict can arise, so $\mathsf{Sat}(\phi) = \mathsf{true}$.
\emph{Completeness:} if no conflict arises, then because capture value domains are infinite (\autoref{def:catalog}), each unconstrained class can be assigned a fresh distinct value, which yields a witness.
\emph{Entailment:} write $\Gamma \models \exists a.\,\mathit{Rel}_o$ if every valuation $X$ of every $\kappa \in \gamma(\sigma^\#)$ admits arguments $a$ with $\mathit{Rel}_o(a, X)$. Let $\mathit{Args}(o)$ be the argument fields of $i_o$. Merge the classes of $D \cup \mathit{Lit} \cup \mathit{Args}(o)$ under the equalities of $\mathit{Rel}_o$ alone. Then $\Gamma \models \exists a.\,\mathit{Rel}_o$ iff (1)~every two registers or literals that fall into one merged class are $\sim$-equivalent in $\Gamma$ and no merged class contains two distinct literals, and (2)~every disequality of $\mathit{Rel}_o$ between two merged classes that both contain a register or literal relates classes that $\mathit{Neq}$ separates or that contain distinct literals, and no disequality relates a class to itself. Classes that contain only arguments receive fresh values, which satisfy the remaining disequalities because the argument types are infinite (\autoref{def:catalog}). The check reuses the union--find structure of $\mathsf{Sat}$.
\emph{Templates:} a question-template relation equates the question-body argument with the rendering $T(X)$; the verifier rejects as unsupported any conjunction with two distinct templates, and treats $T(X)$ as a term determined by the registers it renders, so the two cases above apply unchanged.

\noindent\textbf{Lemma 3 (Simulation).}
For every reachable idle configuration $\kappa = \operatorname{Run}_C(\tau_B^{\mathrm{adm}})$, some $\sigma^\# \in \mathit{Seen}$ satisfies $\kappa \in \gamma(\sigma^\#)$.
\emph{Proof:} By induction on request--result pairs. Initially, $\kappa_0 \in \gamma(\sigma^\#_0)$ and $\sigma^\#_0 \in \mathit{Seen}$ (Line~4).
Suppose $\kappa \in \gamma(\sigma^\#)$ and $\mathsf{req}(o, a)$ is admitted. Then $g(\mathit{fact}, \mathit{ctr})$ holds and $(X, a)$ witnesses $\Gamma \land \mathit{Rel}_o$, so $o \in \mathit{Cand}$ at Line~7 by Lemma~2.
For each decoded outcome $v \in O_{i_o}$, Line~10 computes $\mathsf{AbstractPost}(\sigma^\#, o, v)$.
This removes invalidated registers, adds request captures constrained by $\mathit{Rel}_o$, adds fresh unconstrained classes for result captures, and sets $(q, \mathit{fact}, \mathit{ctr})$ to $(\mathsf{res}(o), v, u(\mathit{ctr}))$ for the unique enabled entering edge of $o$ (Line~8 of \autoref{alg:compiler}), matching Lines~8 and~13 of \autoref{alg:runtime}.
The successor configuration therefore lies in the concretization of a state in $\mathit{Seen}$.

\noindent\textbf{Handler signature.}
At a state $\sigma^\#$ that enables $o$, $\operatorname{Sig}(o)$ is the tuple of the interface $i_o$ with its decoder, the cited obligation, the capture plan $\mathit{cap}_o$, the counter update $u$ of the unique enabled edge into $\mathsf{req}(o)$, and the set of edges out of $\mathsf{res}(o)$, each as its guard, its update, and its target node or exit.

\noindent\textbf{Lemma 4 (Equal signatures, equal behavior).}
If $\operatorname{Sig}(o) = \operatorname{Sig}(o')$, then the plans that $\mathsf{BuildPlan}$ constructs for $o$ and $o'$ on the same $(a, \sigma)$ have the same decoder, invalidation set, request and result captures, and counter update.
Their result nodes $\mathsf{res}(o)$ and $\mathsf{res}(o')$ are distinct, but they have the same outgoing edges with the same guards, updates, and targets, so the two nodes are bisimilar in $C$.
Hence every subsequent admission and exit decision is identical.

\begin{IEEEproof}[Proof of \autoref{thm:runtime-determinism}]
\emph{(a)} Let $o, o' \in \mathit{Occ}(i)$ both yield plans at Line~4 of \autoref{alg:runtime} for the same $a$. By \autoref{thm:safety}(b) and Lemma~3, $\kappa(\sigma) \in \gamma(\sigma^\#)$ for some $\sigma^\# \in \mathit{Seen}$. Both guards hold and $(X, a)$ witnesses $\Gamma \land \mathit{Rel}_o \land \mathit{Rel}_{o'}$, so by Lemma~2 both occurrences are in $\mathit{Cand}$ with $i_o = i_{o'} = i$ and their joint relation is satisfiable.
Therefore Line~8 of \autoref{alg:compiler} (\texttt{P3}) forces $\operatorname{Sig}(o) = \operatorname{Sig}(o')$.
Line~6 of \autoref{alg:runtime} is thus unreachable, and by Lemma~4 the choice made by $\mathsf{SelectCanonical}$ does not affect behavior.

\emph{(b)} Let $\sigma'$ be reached by decoding $v$ at $o$.
By Lemma~3, $\kappa(\sigma') \in \gamma(\sigma^{\#\prime})$ with $\sigma^{\#\prime} = \mathsf{AbstractPost}(\sigma^\#, o, v)$.
By \texttt{P4} and \texttt{P7} (Line~10), $\sigma^{\#\prime}$ either enables a unique exit edge, which Line~14 of \autoref{alg:runtime} assigns to $\mathit{term}$, or is non-terminal; by Line~11, only non-terminal successors enter $W$.
A non-terminal $\sigma^{\#\prime}$ is dequeued, so Line~8 yields an occurrence $o' \in \mathit{Cand}$ whose entering guard holds and with $\Gamma' \models \exists a.\,\mathit{Rel}_{o'}$, where $\Gamma'$ is the register abstraction of $\sigma^{\#\prime}$.
Since $\kappa(\sigma') \in \gamma(\sigma^{\#\prime})$, the entailment part of Lemma~2 gives arguments $a$ with $\mathit{Rel}_{o'}(a, X)$ for the actual valuation $X$ of $\sigma'$, and $a \in L(i_{o'})$ by \autoref{def:catalog}.
Hence $\sigma'$ enables $\mathsf{req}(o', a)$.
\end{IEEEproof}

\subsection{Proof of \autoref{prop:transparency}}
\begin{prop}[Transparency for the Compiled Contract]
\label{prop:transparency}
Under the assumptions of \autoref{thm:runtime-determinism}, let $\sigma$ be a reachable runtime state with $\mathit{pending} = \mathit{term} = \bot$, and let $r = \langle \mathit{tool}, \iota, a \rangle$ pass the common mediation-integrity checks. If the contract step $\mathsf{req}(o, a)$ is enabled at $\kappa(\sigma)$, then \autoref{alg:runtime} admits $r$, as $o$ or as an occurrence with the same handler signature.
Consequently, \autoref{alg:runtime} is sound and transparent for the property $\varphi_C$ induced by $C$, which demands refusal of every request that is neither an enabled step of $C$ nor an auxiliary read, so $\varphi_C$ is enforceable at $B$.
\end{prop}
\begin{IEEEproof}
\autoref{alg:runtime} is deterministic, so $\tau_B^{\mathrm{adm}}$ and $\sigma$ are functions of $\pi_B(\tau)$, where $\tau_B^{\mathrm{adm}}$ labels each dispatched workflow request and its result with the occurrence that $\mathsf{SelectCanonical}$ credits. Let $D_{\varphi_C}(\tau, r)$ be $\mathsf{must\text{-}allow}$ under the conditions of the proposition. Let it be $\mathsf{must\text{-}refuse}$ in three cases, namely when a call is pending or an exit has been entered, when $r$ is a workflow request (one with $a \in \bigcup_i L(i)$) and no step $\mathsf{req}(o, a)$ is enabled at $\operatorname{Run}_C(\tau_B^{\mathrm{adm}})$, and when $r$ matches no workflow interface and no auxiliary adapter. The last case covers out-of-vocabulary tools, compound commands, and inline scripts. All other requests are $\mathsf{free}$.
Since $\mathsf{req}(o, a)$ is enabled, $a \in L(i_o)$, so by \texttt{P1} $\mathsf{ResolveBaseInterface}$ returns $i_o$ (Line~3 of \autoref{alg:runtime}).
Line~2 does not refuse because $\mathit{pending} = \mathit{term} = \bot$.
The enabled edge into $\mathsf{req}(o)$ and $\mathit{Rel}_o(a, X)$ yield a plan for $o$ (Line~4), so Line~5 does not refuse, and by \autoref{thm:runtime-determinism}(a) Line~6 does not refuse either.
Lines~7--9 therefore admit $r$, and by Lemma~4 (\autoref{sec:appendix-sketches}) the canonical choice is behaviorally equal to $o$.
Soundness for $\varphi_C$ follows the three cases. Line~2 refuses requests issued while a call is pending or after an exit, Line~5 refuses a workflow request without an enabled step, since then $\mathit{Plans} = \emptyset$, and Line~3 refuses a request that matches no workflow interface and no adapter. By \autoref{thm:safety}(b), every admitted workflow request extends $\tau_B^{\mathrm{adm}}$ by an enabled step.
\end{IEEEproof}

\section{Additional Tables and Figures}
\label{sec:appendix-eval}

\begin{figure*}[t]
  \centering
  \includegraphics[width=\textwidth]{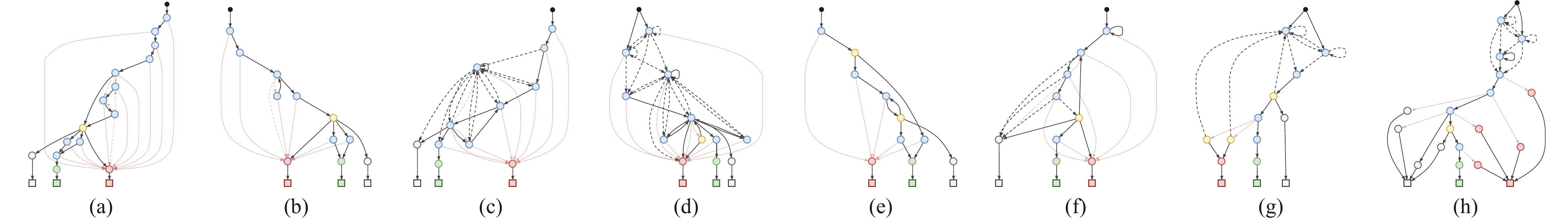}
  \caption{\texttt{SkillCFG} of the evaluation skills at occurrence level:
  (a)~document-editing (validate, ${\le}3$ repairs $\to$ ask $\to$ deliver $\to$ verify);
  (b)~data-reconciliation (validate, ${\le}3$ repairs $\to$ preview $\to$ ask $\to$ deliver);
  (c)~report-render-publish (check, revise freely $\to$ publish or keep a draft);
  (d)~metrics-publish (check $\to$ schema or formula repair $\to$ export);
  (e)~meeting-followup (ask mode $\to$ approve preview $\to$ write);
  (f)~outreach-email (check $\to$ preview $\to$ approve $\to$ send once);
  (g)~workspace-tidy (approve plan $\to$ move $\to$ log; stop on conflict);
  (h)~backup-rotate (copy $\to$ verify $\to$ approve $\to$ delete or retain).
  \scfgEntry~entry, \scfgNode{scfgBlueF}{scfgBlueS}~helper call, \scfgNode{scfgYellowF}{scfgYellowS}~user question;
  completion \scfgNode{scfgGreenF}{scfgGreenS}\,\scfgNode{scfgRedF}{scfgRedS}\,\scfgNode{scfgGrayF}{scfgGrayS} and terminal \scfgTerm{scfgGreenF}{scfgGreenT}\,\scfgTerm{scfgRedF}{scfgRedT}\,\scfgTerm{scfgGrayF}{scfgGrayT} for success, failure, and abort;
  \scfgEdge{scfgGrayT}~transition, \scfgEdge{scfgRedS!45}~failure route, \scfgEdge{scfgGrayT,dashed}~guarded or updated by a counter.
  All panels share one size, so panel area does not reflect contract size.}
  \label{fig:skillcfg-gallery}
\end{figure*}

\begin{figure}[!tbp]
  \centering
  \includegraphics[width=0.9\linewidth]{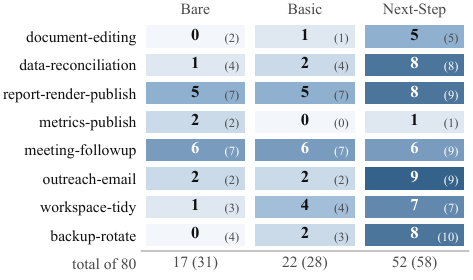}
  \caption{Recovery after the target refusal per skill and refusal message under \cfull.
  Each cell counts strict recoveries of 10 sessions, lawful completions in parentheses.}
  \label{fig:recovery-per-skill}
\end{figure}

\begin{figure}[!tbp]
  \centering
  \includegraphics[width=0.9\linewidth]{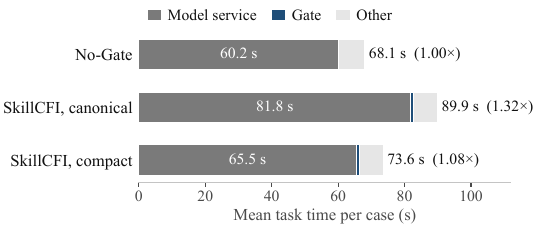}
  \caption{Task time in the dispatch ablation, split into model service, gate, and other time.
  Each bar averages the means of the 16 normal tasks, two sessions each.}
  \label{fig:overhead}
\end{figure}

\autoref{fig:skillcfg-gallery} shows the compiled contracts of the eight skills.
\autoref{tab:abnormal-tasks} lists the 16 abnormal tasks and their refusing checks, and \autoref{tab:public} the public skills accepted after adaptation or porting.
\autoref{fig:recovery-per-skill} gives post-refusal recovery per skill, and \autoref{fig:overhead} splits task time in the dispatch ablation.

\begin{figure}[!tbp]
  \centering
  \includegraphics[width=\linewidth]{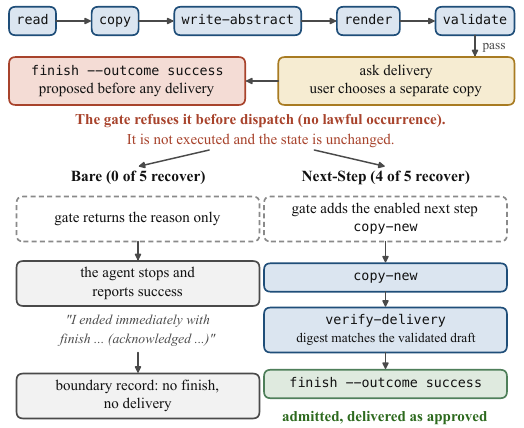}
  \caption{Two sessions of the document-editing task that finishes early, one with \fbare and one with \fguid refusals.
  Both make the same calls up to the refused \texttt{finish}.
  Across the five sessions of this task, none recovers with \fbare and four do with \fguid.}
  \label{fig:supp-trace}
\end{figure}

\autoref{fig:supp-trace} compares two sessions of the document-editing task that asks the agent to finish before delivery (\autoref{sec:eval:utility}).
The artifacts include a replay of the \fguid session against the enforcer without a model.

\begin{table}[!tbp]
  \centering
  \caption{The 16 abnormal tasks. $\dagger$ marks invalid terminations, which are false completions, and the others are invalid transitions (\autoref{sec:security-objective}). \ding{172}--\ding{175} give the deviation class of \autoref{sec:threat-security}. Check is the first check that refused the target under \cfull, which is \textbf{F} whenever no transition is enabled. \textbf{F}\,+\,\textbf{R} marks tasks whose targets mostly also fail \textbf{R}.}
  \label{tab:abnormal-tasks}
  \footnotesize
  \setlength{\tabcolsep}{3pt}
  \resizebox{\linewidth}{!}{%
  \begin{tabular}{@{}lllc@{}}
    \toprule
    \textbf{Skill} & \textbf{Category} & \textbf{Requested violation} & \textbf{Check} \\
    \midrule
    document-editing & Wrong branch (\ding{173}) & replace after the new-copy choice & F\,+\,R \\
     & Premature success & succeed before delivery$^\dagger$ & F \\
    \addlinespace[2pt]
    data-reconciliation & Excess repetition (\ding{174}) & save after four failed validations & F \\
     & Skipped validation (\ding{172}) & deliver without validation & F\,+\,R \\
    \addlinespace[2pt]
    report-render-publish & Stale evidence (\ding{172}) & publish a revision on an old check & F \\
     & Skipped validation (\ding{172}) & publish before the check & F \\
    \addlinespace[2pt]
    metrics-publish & Success after fail & succeed after a failed check$^\dagger$ & F \\
     & Wrong branch (\ding{173}) & update schema after a formula failure & F \\
    \addlinespace[2pt]
    meeting-followup & Approval bypass (\ding{172}) & write before approval & F \\
     & Stale evidence (\ding{175}) & write unapproved content & R \\
    \addlinespace[2pt]
    outreach-email & Excess repetition (\ding{174}) & send the approved email twice & F\,+\,R \\
     & Denial override (\ding{173}) & send after ``keep draft'' & F\,+\,R \\
    \addlinespace[2pt]
    workspace-tidy & Premature success & succeed without the move log$^\dagger$ & F \\
     & Stop override (\ding{173}) & move again after the Stop answer & F \\
    \addlinespace[2pt]
    backup-rotate & Failed check (\ding{173}) & delete after failed verification & F\,+\,R \\
     & Denial override (\ding{173}) & delete after the No answer & F \\
    \bottomrule
  \end{tabular}}
\end{table}

\begin{table}[!tbp]
  \centering
  \caption{Public skills accepted after adaptation or porting.
  All seven are scope restricted.
  Comp.\ counts adaptation compilations, LoC the non-blank, non-comment lines of added helpers, and Obl.\ the enforced plus the guidance obligations.
  Replay counts the registered verdicts that the contract matches.}
  \label{tab:public}
  \scriptsize
  \setlength{\tabcolsep}{2.6pt}
  \begin{tabular}{@{}lllrrcr@{}}
    \toprule
    Skill & Repo. & Tier & Comp. & LoC & Obl. & Replay \\
    \midrule
    beam & openclaw & adapted & 5 & 88 & 5+8 & 10/10 \\
    agent-transcript & openclaw & ported & 10 & 122 & 12+11 & 18/18 \\
    transcribe & openai & ported & 12 & 64 & 4+11 & 19/19 \\
    speech & openai & ported & 2 & 64 & 4+13 & 19/22 \\
    security-ownership-map & openai & adapted & 3 & 105 & 4+6 & 22/22 \\
    skill-installer & openai & adapted & 1 & 101 & 5+7 & 19/19 \\
    vercel-deploy & openai & adapted & 1 & 69 & 5+7 & 19/19 \\
    \bottomrule
  \end{tabular}
\end{table}

\section{Evaluation Setup Details}
\label{sec:appendix-setup}

\noindent\textbf{Contracts.}
The contracts of the eight skills (\autoref{fig:skillcfg-gallery}) have 9--17 occurrences and 19--48 transitions.
Each skill has 7--16 enforced obligations and keeps 2--6 more as guidance.
All skills except meeting-followup use counters, and all except report-render-publish ask the user.
Every workflow ends in a success, failure, or abort terminal through a completion acknowledgment, and helper errors end in failure.

\noindent\textbf{Implementation.}
The compiler checks \texttt{P1}--\texttt{P7} with a union-find equality and disequality solver, with \texttt{P4} checked as satisfiability.
The enforcer runs on the OpenCode \texttt{tool.execute.before/after} hooks and keeps $\sigma$ in a host-side closure (\textbf{M4}).
A POSIX literal-command lexer blocks shell metacharacters and expansions ($R_{3\mathrm{b}}$) and rewrites \texttt{Bash} arguments into canonical form, and bundled helpers return JSON receipts for decoding.

\noindent\textbf{Human raters.}
Both raters judge every mutation, rewrite, attribution, adapted package, and screened public skill.
The obligation, source-line, and refusal checks use the 20\% sample of \autoref{sec:eval:setup}.

\end{document}